\documentclass[aps,pra,preprint,floatfix]{revtex4-1}

\usepackage{amsmath}
\usepackage{graphicx}
\usepackage{subfigure}
\usepackage{amsfonts}
\usepackage{sidecap}
\usepackage{natbib}
\usepackage{url}

\newcommand{\bx}{\mathbf{x}}
\newcommand{\psisq}{\left| \Psi \right|^{2}}
\newcommand{\dbx}{\,\mathrm{d}\mathbf{x}}
\newcommand{\dx}{\,\mathrm{d}x}
\newcommand{\dy}{\,\mathrm{d}y}
\newcommand{\dt}{\,\mathrm{d}t}
\newcommand{\rhobar}{\overline{\rho}}
\newcommand{\psibar}{\overline{\psisq}}
\newcommand{\cgh}{\overline{H}}
\newcommand{\lang}{\left\langle}
\newcommand{\rang}{\right\rangle}

\begin{document}

\title{Timescales for dynamical relaxation to the Born rule}
\date{\today}
\author{M.D.~Towler}
\author{N.J.~Russell}
\affiliation{Theory of Condensed Matter Group, Cavendish Laboratory, University of Cambridge, J.J.~Thomson Avenue, Cambridge CB3 OHE, U.K.}
\author{Antony Valentini}
\affiliation{Department of Physics and Astronomy, Clemson University, 303 Kinard Laboratory, Clemson SC 29631-0978, U.S.A.}

\begin{abstract}

\noindent We illustrate through explicit numerical calculations how the
Born-rule probability densities of non-relativistic quantum mechanics emerge
naturally from the particle dynamics of de Broglie-Bohm pilot-wave theory. The
time evolution of a particle distribution initially \emph{not} equal to the
absolute square of the wave function is calculated for a particle in a
two-dimensional infinite potential square well. Under the de Broglie-Bohm
ontology, the box contains an objectively-existing `pilot wave' which guides the
electron trajectory, and this is represented mathematically by a Schr\"odinger
wave function composed of a finite out-of-phase superposition of \(M\) energy
eigenstates (with $M$ ranging from 4 to 64). The electron density distributions
are found to evolve naturally into the Born-rule ones and stay there; in analogy
with the classical case this represents a decay to `quantum equilibrium'. The
proximity to equilibrium is characterized by the coarse-grained subquantum
\emph{H}-function which is found to decrease roughly exponentially towards
zero over the course of time. The timescale \(\tau\) for this relaxation is
calculated for various values of \(M\) and the  coarse-graining length
\(\varepsilon\). Its dependence on $M$ is found to disagree with an earlier
theoretical prediction. A power law, \(\tau \propto M^{-1}\), is found to be
fairly robust for all coarse-graining lengths and, although a weak dependence of
\(\tau\) on \(\varepsilon\) is observed, it does not appear to follow any
straightforward scaling. A theoretical analysis is presented to explain these
results. This improvement in our understanding of timescales for relaxation to
quantum equilibrium is likely to be of use in the development of models of
relaxation in the early universe, with a view to constraining possible
violations of the Born rule in inflationary cosmology.

\end{abstract}

\maketitle

\newpage

  \section{Introduction}
  \label{sec:introduction}

The Born rule is the fundamental connection between the mathematical formalism
of quantum theory and the results of experiments. It states that if an
observable corresponding to a Hermitian operator ${\hat A}$ is measured in a
system with pure quantum state $|\Psi\rangle$,  the probability of an eigenvalue
$\lambda_i$ will equal $\lang\Psi|{\hat P}_i|\Psi\rang$, where ${\hat P}_i$ is
the projection onto the eigenspace of ${\hat A}$ corresponding to $\lambda_i$.
For the case of measurement of the position ${\bf x}$ of -- say -- an electron
in a box, the probability density at time $t$ for finding the electron at ${\bf
x}$ is $\rho({\bf x},t) = |\Psi({\bf x},t)|^2$. The Born rule is normally
presented as a postulate, though attempts to derive it from more fundamental
principles have a long history. There has, for example, been much recent work on
deriving the Born rule within the framework of the many-worlds interpretation of
quantum mechanics, but such derivations remain
controversial~\cite{manyworldsbook}. According to a recently-published (2008)
encyclopedia of quantum mechanics ``\emph{the conclusion seems to be that no
generally accepted derivation of the Born rule has been given to date, but this
does not imply that such a derivation is impossible in
principle}''~\cite{compendium08}. 

Born's 1926 paper~\cite{born26}, and  Heisenberg's introduction of the
uncertainty relations the following year~\cite{heisenberg27}, were instrumental
in popularizing the idea that Nature at the quantum level is fundamentally
probabilistic. The idea that $\Psi$ provides a \emph{complete} description of a
single electron strongly suggests that the probabilistic interpretation of
$|\Psi|^2$ expresses an irreducible uncertainty in electron behaviour that is
intrinsic in Nature. It is somewhat ironic therefore - and unknown to most
physicists - that Born's rule emerges quite naturally out of the dynamics of a
\emph{deterministic} process that was first outlined by de Broglie in
1927~\cite{valentini_book}. The process in question can be described by a theory
commonly referred to as the de Broglie-Bohm `pilot-wave' formulation of quantum
mechanics~\cite{debroglie28, bohm52a, *bohm52b, bohm_book, holland_book,
durr_book, riggs_book, valentini_physics_world, gradcourse}. While the theory
has attracted little serious interest ever since it was introduced~\cite{cushing_book,
valentini_book}, there has been a considerable resurgence of activity in this area over
the last fifteen years or so~\cite{tticonf}. One of the reasons for this is that, although the theory is completely consistent with the full
range of predictive-observational data in quantum mechanics, it also permits
\emph{violations} of the Born rule and, at least in principle, this leads to the
possibility of new physics and of experimentally testable
consequences~\cite{valentini91a, *valentini91b, valentini02, valentini07,
valentini10}. 

The reason that de Broglie-Bohm theory can get away with such an apparently
absurd contradiction of one of the basic postulates of quantum theory is that it
assumes orthodox quantum mechanics is \emph{incomplete}, as Einstein always
insisted. It supposes that electrons, for example, are real `particles' with
continuous trajectories and that the Schr\"odinger wave function represents an
objectively-existing `pilot wave' which turns out to influence the motion of the
particles. Since  the particle density $\rho$ and the square of the pilot wave
are logically distinct entities, they can no longer be
\emph{postulated} to be equal to each other. Rather, their identity should be
seen as dynamically generated in the same sense that one usually regards thermal
equilibrium as arising from a process of relaxation based on some underlying
dynamics (though with a dynamics on configuration space rather than phase
space). Since pilot-wave theory features a different set of basic axioms and
conceptual structures, with event-by-event causality and the prospect of making
predictions different from orthodox QM, it is better to think of it as a
different theory, rather than a mere `interpretation' of quantum mechanics. 

In the pilot-wave formulation then, quantum mechanics emerges as the statistical mechanics of the underlying deterministic theory. If the particle distribution obeys the Born rule $\rho = \psisq$, the system is said to be in `quantum equilibrium'. One finds in general that:

1. Non-equilibrium systems naturally tend to become Born-distributed over the course of time, on a coarse-grained level, provided the initial conditions have no fine-grained microstructure~\cite{valentini_thesis, valentini91a, *valentini91b, valentini01, valentini05}. (The latter restriction is similar to that required in classical statistical mechanics. An assumption about initial conditions is of course required in any time-reversal invariant theory in order to demonstrate relaxation~\cite{valentini01,valentini05}.)

2. Once in quantum equilibrium a system will remain in equilibrium thereafter, as was originally noted by de Broglie~\cite{debroglie28} (this property is sometimes referred to as `equivariance').

These two observations - along with a description of how these
statements about the objective makeup of the system might be translated into
statements about measurement - can be said to `explain' or derive the
Born rule. Given the common general viewpoint referred to in the first
paragraph, many physicists might consider this surprising.

In this work, we present a numerical analysis of the \emph{timescale} for the
relaxation of non-equilibrium distributions of particles to Born-rule quantum
equilibrium using pilot-wave dynamics; the approach to equilibrium is monitored
by computing the coarse-grained subquantum \emph{H}-function (see
section~\ref{sec:equilibrium} and Ref.~\onlinecite{valentini91a,
*valentini91b}). The results we obtain are for a particle in a 2D infinite
potential square well where the wave function is a finite superposition of
\(M\) eigenfunctions (where, depending on the choice of initial state, $M$
ranges from 4 to 64). The initial particle distribution is deliberately chosen
to be `out of equilibrium' by giving it the same form as the absolute square of
the ground-state wave function, that is, \(\rho = 4 / \pi^2 \sin^{2}x \sin^{2}
y\).  This system - with fixed $M$ - has been studied before in this context by
Valentini and Westman~\cite{valentini05} and by Colin and
Struyve~\cite{colin09}, but here we go further. Our recent development of a new
and much faster computer code~\cite{louis} allows us to study systems with many
more modes. The timescale \(\tau\) for relaxation is studied as a function of
the number of modes $M$ (and, in consequence, of the number of nodal points in
the wave function) and as a function of the coarse-graining
length~$\varepsilon$.  The dependence of the relaxation timescale on these two
quantities is compared to theoretical predictions. 

It is intended that calculations such as these will provide a next step towards
a detailed understanding of relaxation to quantum equilibrium in the early
universe, with a view to constraining possible non-equilibrium effects in
cosmology. In Ref.~\onlinecite{valentini10}  it was shown, in the context of
inflationary cosmology, that corrections to the Born rule in the early universe
would in general have potentially observable consequences for the cosmic
microwave background (CMB). This is because, according to inflationary theory,
the primordial perturbations that are currently imprinted on the CMB were
generated at early times by quantum vacuum fluctuations whose spectrum is
conventionally determined by the Born rule. To make detailed predictions for
possible anomalies in the CMB, however, requires a precise understanding of how
fast relaxation would occur in, for example, a pre-inflationary era (as
discussed in section IV-A of Ref.~\onlinecite{valentini10}). It may be hoped
that numerical studies, such as those reported in this paper, will reveal how
the relaxation timescale depends on general features of the quantum state such
as the number $M$ of modes in a superposition. The results could then be
applied in future work to specific cosmological models.

\subsection{Pilot-wave dynamics}
\label{sec:dynamics}

The basic ideas of de Broglie-Bohm pilot-wave theory may be simply understood in
a non-relativistic context~\footnote{In the high-energy domain, pilot-wave
theory for bosons usually takes the form of a field theory, while for fermions
the best model invokes the Dirac sea. These models have a fundamental preferred
rest frame, with an effective Lorentz invariance emerging only in equilibrium.
For recent progress see in particular Refs.~\onlinecite{colin03, colinstruyve07,
struyve10}.}. It is a non-local hidden-variables theory, that is, the theory
contains some variables that distinguish the individual members of an ensemble
that in orthodox QM would be considered identical since they all have the same
wave function. These variables are supposed to be ultimately responsible for the
apparently random nature of - for example - position measurements on the
system.  If, as required by some interpretations, one were to suppose both
that a complete description of the system is afforded by $\Psi$ and that $\Psi$
has an objective, physical existence, one might conclude from the results of
measurements that Nature is \emph{intrinsically} probabilistic or random. In
pilot-wave theory, by contrast, one supplements the wave function description
with `hidden variables' by postulating the existence of particles with definite
positions, in addition to the wave. These particles then follow deterministic
trajectories (the nature of which can be deduced) and the observed randomness is
then understood to be a consequence merely of our ignorance of the initial
conditions, that is, the starting positions of the particles.

How does an individual quantum system evolve in time? The pilot wave  evolves at
all times according to the usual time-dependent Schr\"odinger equation 
\begin{eqnarray*}
i\hbar
\frac{\partial\Psi}{\partial t} = \sum_{i=1}^M -
\frac{\hbar^2}{2m_i}\nabla_i^2\Psi + V\Psi .
\end{eqnarray*}
As normally understood the evolving quantum system behaves like a `probability
fluid' of density $|\Psi|^2 = \Psi\Psi^*$ with an associated time-dependent
quantum probability current, defined in the usual manner as ${\bf j} =
\frac{\hbar}{m} \mathrm{Im} (\Psi^*{\bf \nabla}\Psi)$. In pilot-wave theory, the
particles have a continuous objective existence, with trajectories that follow
the streamlines of the current. Thus their velocity is given by the current
divided by the density, that is, by 
\begin{eqnarray*}
{\bf v} = \frac{\hbar}{m} \mathrm{Im} \nabla \ln \Psi .
\end{eqnarray*}
Using the complex polar form of the wave function $\Psi = |\Psi|\exp[iS/\hbar]$,
we can recover the (locally defined) phase $S({\bf x}_1, \ldots,{\bf x}_N,t)$ of
the wave by the expression  $S=\hbar \mathrm{Im} \ln \Psi$. The \emph{de Broglie guidance equation} for the trajectories ${\bf x}_i(t)$ may then be written as 
\begin{eqnarray}  
\label{eqn:debgeq}
\frac{d{\bf x}_i}{dt} = \frac{\nabla_i S}{m_i}   
\end{eqnarray}
If, for an ensemble of particles with the same wave function, the initial positions have a Born-rule distribution, then (by construction) the law of motion of Eqn.~\ref{eqn:debgeq} implies that the particle positions will have a Born-rule distribution at all times.

If desired, one may take the first time derivative to write the equation of motion in second-order form, 
\begin{eqnarray}
\label{eqn:bgeq}
m_{i}\mathbf{\ddot{x}}_{i}=-\mathbf{\nabla}_{i}(V+Q) , 
\end{eqnarray}
where the \emph{quantum potential}
$Q=-\sum_{i}\frac{\hbar^{2}}{2m_{i}}\frac{\nabla_{i}^{2}\left\vert
\Psi\right\vert }{\left\vert \Psi\right\vert }$. In this approach, the system
acts as if there were a `quantum force' $-\nabla_i Q$ acting on the particles in
addition to the classical force $-\nabla_i V$. This second-order approach with a
law of motion given by Eqn.~\ref{eqn:bgeq} was proposed by Bohm in 1952. It may
be referred to as `Bohm's dynamics' in order to distinguish it from `de
Broglie's dynamics' based on Eqn.~\ref{eqn:debgeq} (which was proposed by de
Broglie in 1927). For de Broglie, ${\bf p} = \nabla S$ is the law of motion; for
Bohm -- at least the Bohm who wrote the 1952 papers -- it is an initial
condition which can be dispensed with (clearly if we integrate the second-order
formula we only recover de Broglie's equation up to some constant and this must
be fixed for each trajectory by some boundary condition,  such as that implied
by de Broglie's equation for some time $t_0$). Thus, in principle, Bohm's
dynamics encompasses what one might call `extended nonequilibrium' where ${\bf
p} \neq \nabla S$ in addition to $\rho \neq |\Psi|^2$. Recent
work~\cite{bohm_instability} suggests that this `extended nonequilibrium' is
unstable and does not relax in general; if this is correct then it may be argued
that Bohm's second-order dynamics is untenable as a fundamental theory as there
would be no reason to expect equilibrium in the universe today, and that de
Broglie's dynamics is in fact the fundamental formulation of pilot-wave theory.

Some additional relevant observations:

(1) The form of the guidance equation may be altered, while retaining
consistency with the Born-rule distribution. This can be achieved by adding a
divergence-free term (divided by $|\Psi|^2$) to the right-hand side. Such
alternative velocity fields will not be discussed further here but have been
studied by, for example, Colin and Struyve~\cite{colin09} and Timko and
Vrscay~\cite{timko09}. Note that such alternatives yield an equivalent physics
only in the equilibrium state; away from equilibrium, `subquantum' measurements
would allow one to track the trajectories and so distinguish the true velocity
field~\cite{valentini02}.

(2) Given the wave function for a system, the particle trajectories from any
starting point may be calculated using only the initial position of the
particle, rather than the position and the momentum. This is because the
guidance equation alone gives the particle velocity and consequently the
momentum for any initial position.

\begin{figure*}[t]
  \begin{center}
    \subfigure[(1.5,1.55)]{
    \includegraphics[width=0.3\textwidth]{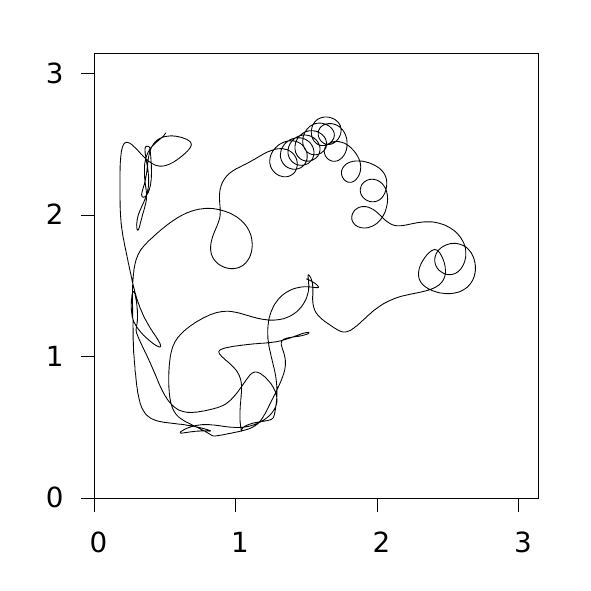}
    \label{fig:traj1}}
    \subfigure[(1.5,1.56)]{
    \includegraphics[width=0.3\textwidth]{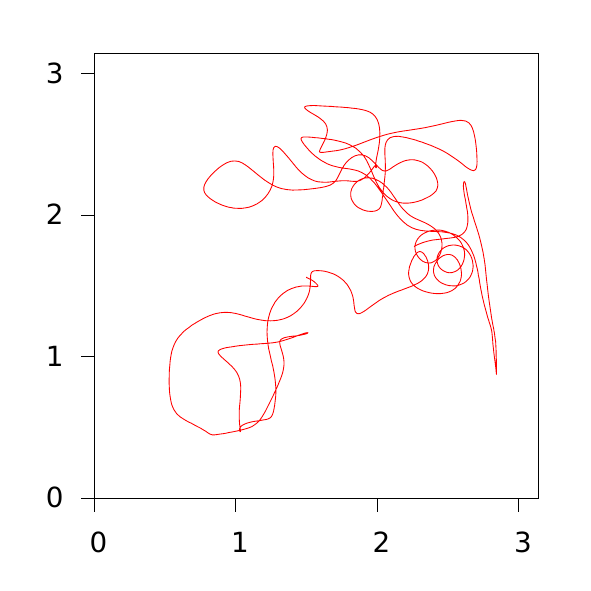}
    \label{fig:traj2}}
    \subfigure[both]{
    \includegraphics[width=0.3\textwidth]{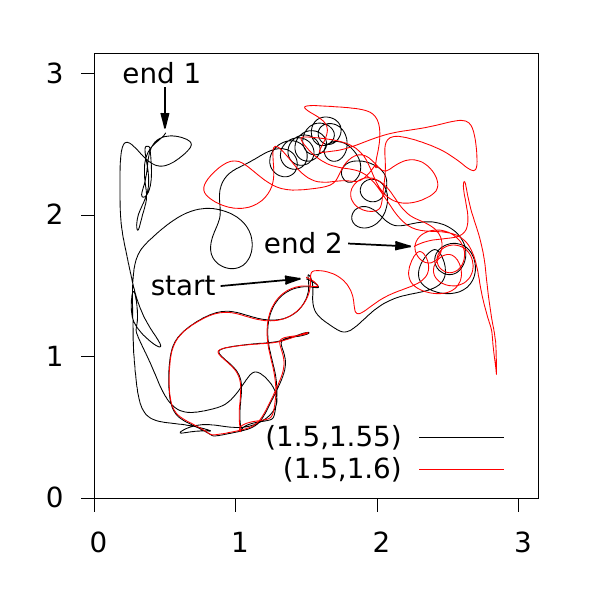}
    \label{fig:traj3}}
    \caption{Comparison of two trajectories with almost identical initial positions, shown individually in
    \ref{fig:traj1} and \ref{fig:traj2} for clarity, then superimposed
    in \ref{fig:traj3}, in a system where the wave function is a superposition of 16 energy eigenfunctions. The particles have been propagated for time
    \(4\pi\) in both cases -- note the rapid divergence.
    \ref{fig:traj1} also shows a good example of particle motion near to
    a node. The trajectory is seen to spiral around a moving nodal point before departing
    from the vicinity of the node (similar behaviour is reported in Ref.~\onlinecite{valentini05}). This behaviour seems to be
    a major driving force behind
    relaxation.}
    \label{fig:divergence}
  \end{center}
\end{figure*}

(3) Particle trajectories tend to be quite erratic, even with simple wave
functions that are superpositions of just a few energy eigenfunctions.
Fig.~\ref{fig:divergence} illustrates the divergence of neighbouring particle
trajectories by showing the paths of two particles with almost identical initial
positions, propagating according to pilot-wave dynamics.

How do numerical simulations demonstrating the Born rule for the actual particle
positions translate into statements about `measurement'?  Ideal measurements of
\emph{position} in pilot-wave theory are usually correct measurements (they
reveal the pre-existing position of the particle - see Ref.~\cite{holland_book},
p.351) and so the Born rule in position space follows immediately if the
particles really are distributed that way. For other kinds of measurements, a
clear derivation of the Born rule may be found in Section 8.3.5 of Holland's
textbook~\cite{holland_book} (noting that Holland assumes the $|\Psi|^2$
distribution of actual particle positions as a \emph{postulate} - see
Ref.~\onlinecite{holland_book}, p.67). The key point is that, in a theory of particles, experimental observations may be reduced to particle positions (dots on screens, apparatus pointer positions, etc.) - where laboratory apparatus is treated as just another system made of particles. As long as the Born rule holds for the joint distribution of positions of all the particles involved (including the particles making up the equipment), then the marginal probability distribution for, say, pointer positions (obtained by integrating out the other degrees of freedom) will necessarily match the predictions of quantum mechanics. In such a case, the distribution of macroscopically-recorded outcomes will be the same as in quantum theory.

\subsection{Quantum equilibrium}
\label{sec:equilibrium}

 To demonstrate equivalence to quantum mechanics, it is usually simply assumed
that the actual distribution of particle positions is already supplied to us
obeying the Born rule \(\rho = \psisq\). In the approach taken here, where we
try to demonstrate why this is so, the Born-rule distribution is considered to
be a special case and the particles are said to be in \emph{quantum equilibrium}
when in this state. The dynamics described in section~\ref{sec:dynamics} can
just as well be used to describe the evolution of non-equilibrium systems,
whereas standard formulations cannot. In general in such studies, the
probability density is found to approach the Born rule distribution over time;
it is said to relax to equilibrium~\cite{valentini91a, valentini05}. This
relaxation is a consequence of the deterministic motion of the particles and is
not an intrinsically stochastic process (further insight into relaxation has been obtained by Bennett using techniques from Lagrangian fluid dynamics~\cite{bennett10}). 

\begin{figure*}[h]
  \begin{center}
    \(\psisq\)\\
    \includegraphics[width=0.3\textwidth]{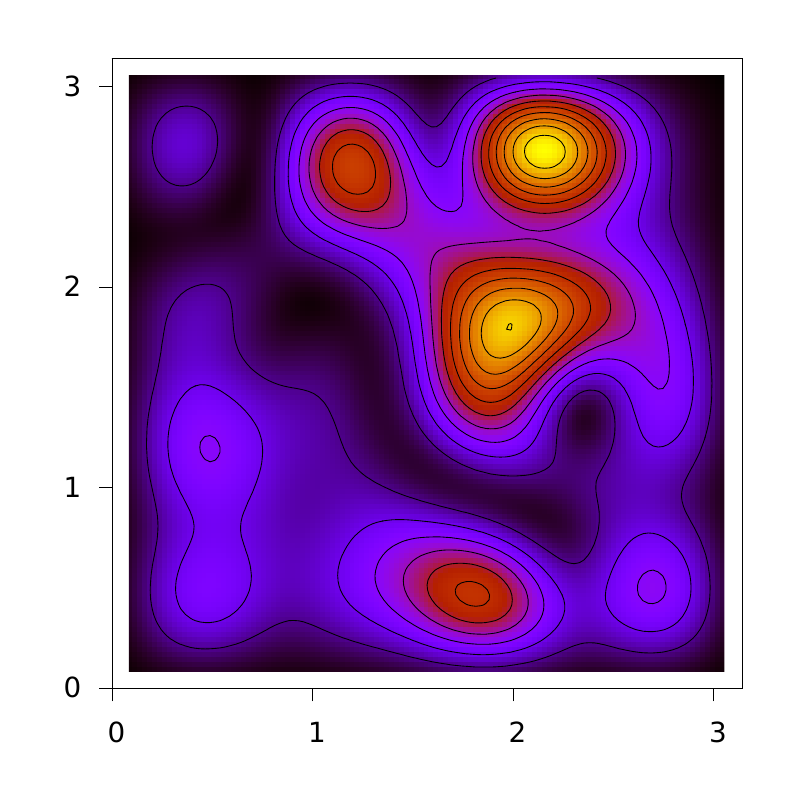}
    \includegraphics[width=0.3\textwidth]{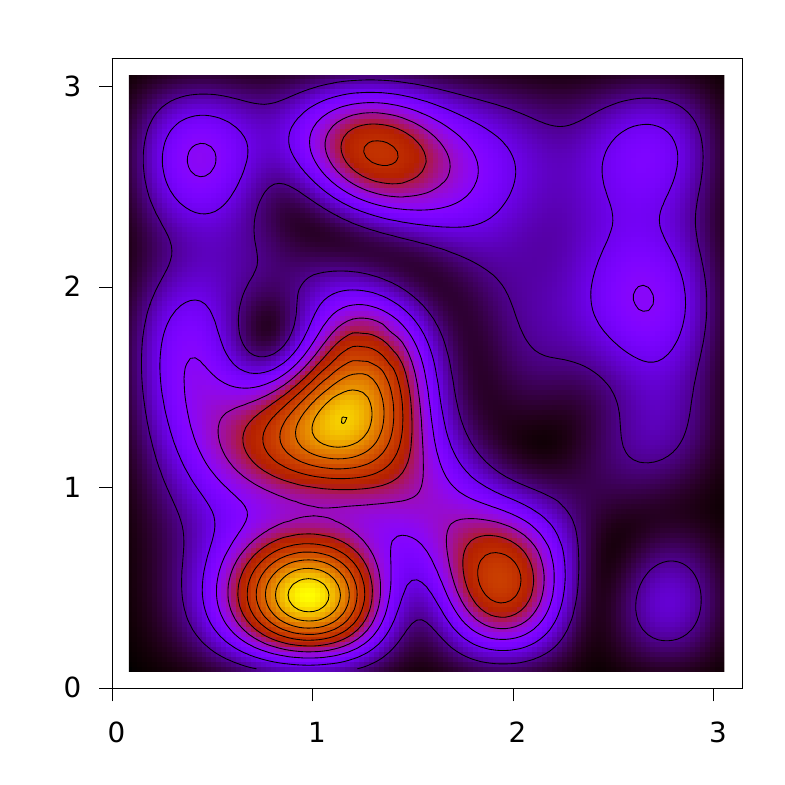}
    \includegraphics[width=0.3\textwidth]{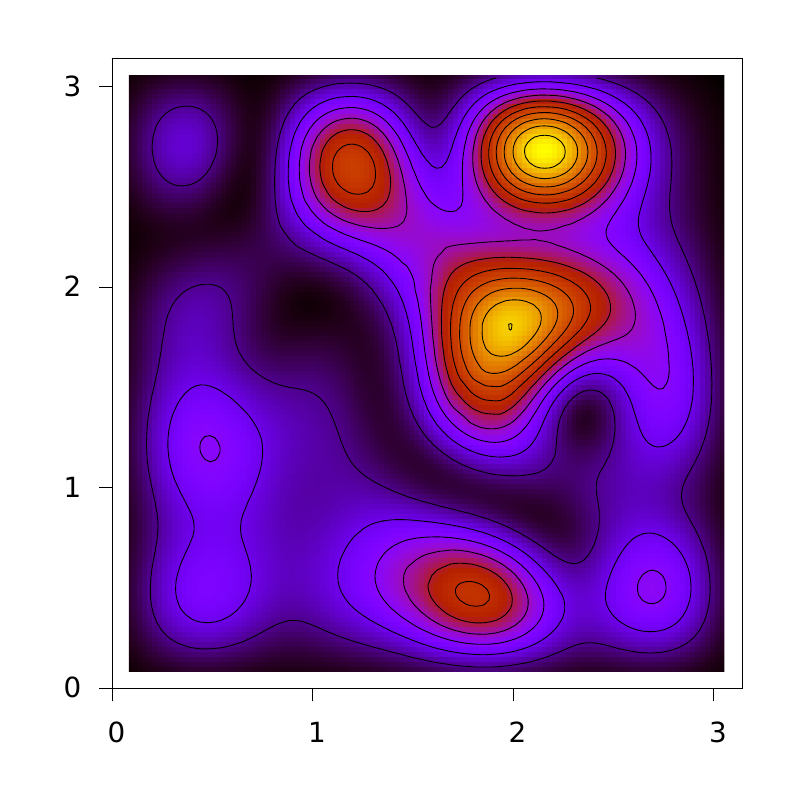}\\
    \(\rho\)\\
    \subfigure[t=0]{
    \includegraphics[width=0.3\textwidth]{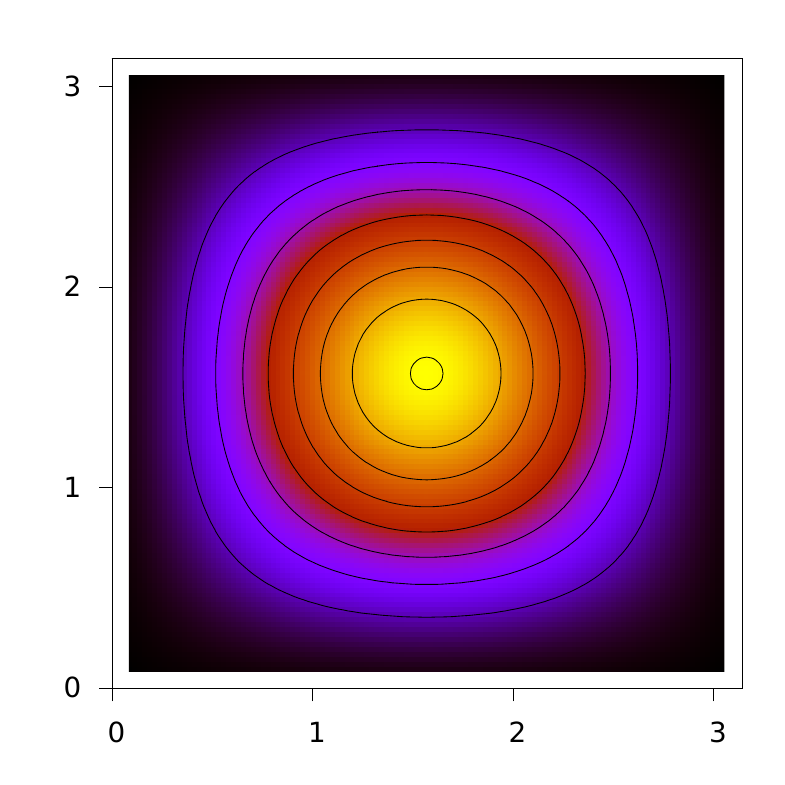}
    }\subfigure[t=2\(\pi\)]{
    \includegraphics[width=0.3\textwidth]{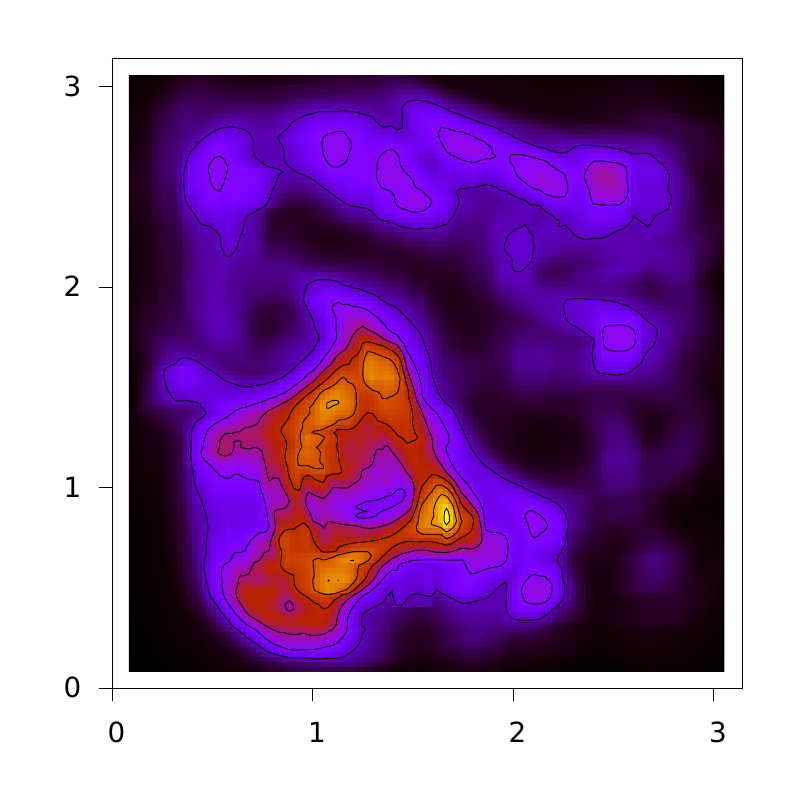}
    }\subfigure[t=4\(\pi\)]{
    \includegraphics[width=0.3\textwidth]{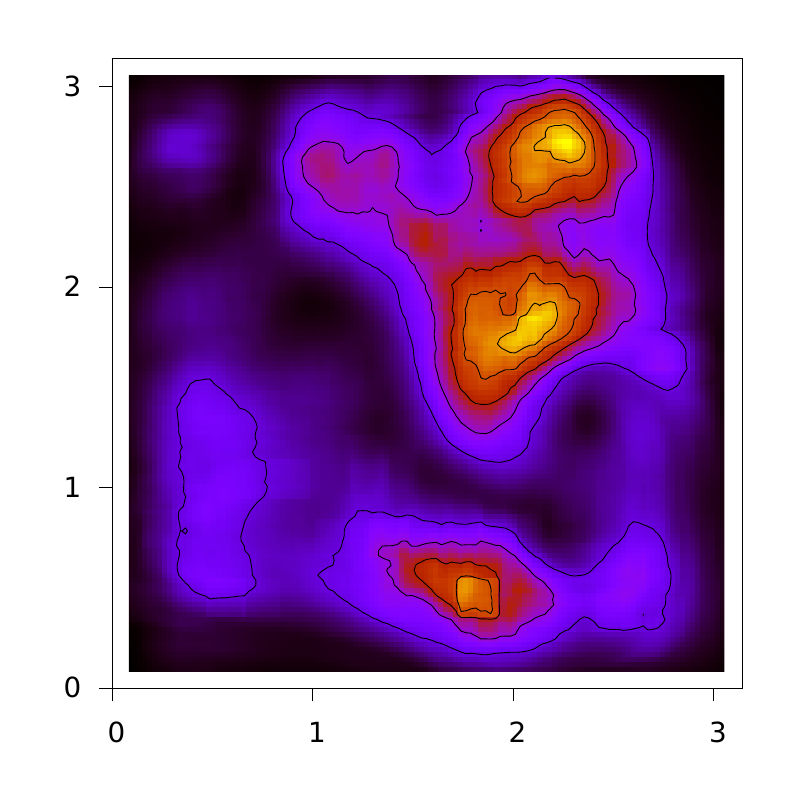}
    }
    \caption{Computational model of the relaxation of an initially
    non-equilibrium distribution, \(\rho = \frac{4}{\pi^2} \sin^{2} x \sin^{2} y\), evolving according to 
    pilot-wave dynamics. The wave function is a superposition of the first sixteen eigenstates
    for a particle in a 2D infinite potential square well.
    The simulation was run for one period of the wavefunction, or \(4\pi\) in
    these units. Even after such a short period, significant relaxation towards equilibrium can be
    observed.  (These results provide an independent confirmation of those first obtained by Valentini and Westman~\cite{valentini05}.)}
    \label{fig:relax1}
  \end{center}
\end{figure*}

Fig.~\ref{fig:relax1} shows the results of a numerical simulation of this
relaxation process. It can be clearly seen that the particle distribution $\rho$
rapidly comes to resemble the (periodically repeating) time-dependent
$|\Psi|^2$. The example chosen - a superposition of sixteen modes, for a particle moving in two spatial dimensions - is identical to that studied by Valentini and Westman~\cite{valentini05}. The results obtained match theirs, thereby providing an important confirmation of the previous results, with an independently written and implemented numerical code~\footnote{It should be noted, however, that in the code used by Valentini and Westman, the signs of the initial phases in the wave function were chosen with the opposite convention to that given in their text (where the latter agrees with Eqn.~\ref{eqn:wavefunc} above). Furthermore, in their plots of density functions, the labels on the $x$- and $y$-axes were inadvertently exchanged.}.

If we are to have any chance of observing new physics associated with quantum
nonequilibrium states~\cite{valentini91a, valentini_thesis, valentini01, valentini02, valentini02b, valentini07, valentini10}, it is important to have a handle on
the timescale of this relaxation. To quantify the proximity of a distribution to
equilibrium, we may use an analogue of the classical 
\emph{H}-function \cite{valentini91a, *valentini91b, valentini05}. This
`subquantum \emph{H}-function' is defined as
\begin{equation}
  \label{eqn:hfunction}
  H = \int \left( \frac{\rho}{\psisq} \right) \ln \left( \frac{\rho}{\psisq}
  \right) \mathrm{d}\Sigma
\end{equation}
where the (weighted) volume element \(\mathrm{d}\Sigma = \psisq \dbx\).

This quantity will be zero if and only if \(\rho = \psisq\) everywhere, and will
be positive otherwise~\footnote{Assuming $\rho$ and $\psisq$ are normalized then $\int (\rho - \psisq)\dbx = 0$. Since $\rho \ln \left(\frac{\rho}{\psisq} \right) \geq \rho - \psisq$ for any value of $\rho$ and $\psisq$, with equality only when $\rho = \psisq$, then it is clear that $H = \int \rho \ln \left( \frac{\rho}{\psisq} \right) \dbx \geq 0$. Thus $H$ is bounded from below by zero.}, making it a useful measure of proximity to
equilibrium. Clearly, $H$ is simply the negative of the relative entropy
of \(\rho\) with respect to \(\psisq\).

A feature of the quantities in this definition is that both the volume
element, \(\mathrm{d}\Sigma\), and the ratio \(f=\frac{\rho}{\psisq}\) are
preserved along trajectories. To show this for \(f\), consider the two
continuity equations:
\begin{align}
  \label{eqn:contrho}
  \frac{\partial \rho}{\partial t} + \nabla \cdot \left( \dot{\bx} \rho
  \right) &= 0   
  \intertext{which follows from the assumption that the actual trajectories follow the velocity field given by Eq.~\ref{eqn:debgeq}, and}
  \label{eqn:contpsi}
  \frac{\partial \psisq}{\partial t} + \nabla \cdot \left( \dot{\bx} \psisq
  \right) &= 0
\end{align}
which follows from the Schr\"odinger equation.\vspace{0.2cm}\\
These two equations can be used to show that the ratio
\(f=\frac{\rho}{\psisq}\) obeys:
\begin{equation}
  \frac{df}{dt} \equiv \frac{\partial f}{\partial t} + \dot{\bx} \cdot \nabla
  f = 0
\end{equation}

Thus \(f\) will be preserved along trajectories. Thus, if the system is
initially in quantum equilibrium, with $f = 1$ everywhere, it will never depart from that state. This can, of course, be seen directly from the fact that $\rho$ and $|\Psi|^2$ obey identical continuity equations: if $\rho$ and $|\Psi|^2$ are initially equal, they will necessarily remain equal at all times, since their time evolutions are determined by the same partial differential equation.

For general (non-equilibrium) initial conditions, the exact value of the \emph{H}-function remains unchanged as the
system evolves. However, if a coarse-graining is applied to \(\rho\) and
\(\psisq\), that is, we replace \(\rho \rightarrow \rhobar\), \(\psisq
\rightarrow \psibar\) where overbars indicate averaging over small
coarse-graining cells, then the coarse-grained \emph{H}-function
\begin{equation}
  \label{eqn:cghfunction}
  \cgh = \int \rhobar \ln \left( \frac{\rhobar}{\; \psibar \;}
  \right) \mathrm{d}{\bf x}
\end{equation}
can be shown to be non-increasing, on the assumption that the initial state
contains no fine-grained microstructure (as in the analogous classical
coarse-graining \emph{H}-theorem)~\cite{valentini91a}. Furthermore, $\cgh$ will
in fact decrease, if the initial velocity field varies with position across the
coarse-graining cells~\cite{valentini_thesis, valentini01}. The decrease of $\cgh$ represents a
relaxation of the system towards equilibrium, and formalizes an analogue of the
intuitive idea of Gibbs: an initial non-equilibrium distribution will tend to
develop fine-grained microstructure and become closer to equilibrium on a
coarse-grained level. Heuristically speaking, this may be thought of in terms
of  two `fluids', with densities $\rho$ and $|\Psi|^2$, that are `stirred' by
the same velocity field, and thereby tend to become indistinguishable when
coarse-graining is applied. 

\begin{figure*}[h]
  \begin{center}
    \subfigure[fine-grained]{
    \includegraphics[width=0.3\textwidth]
    {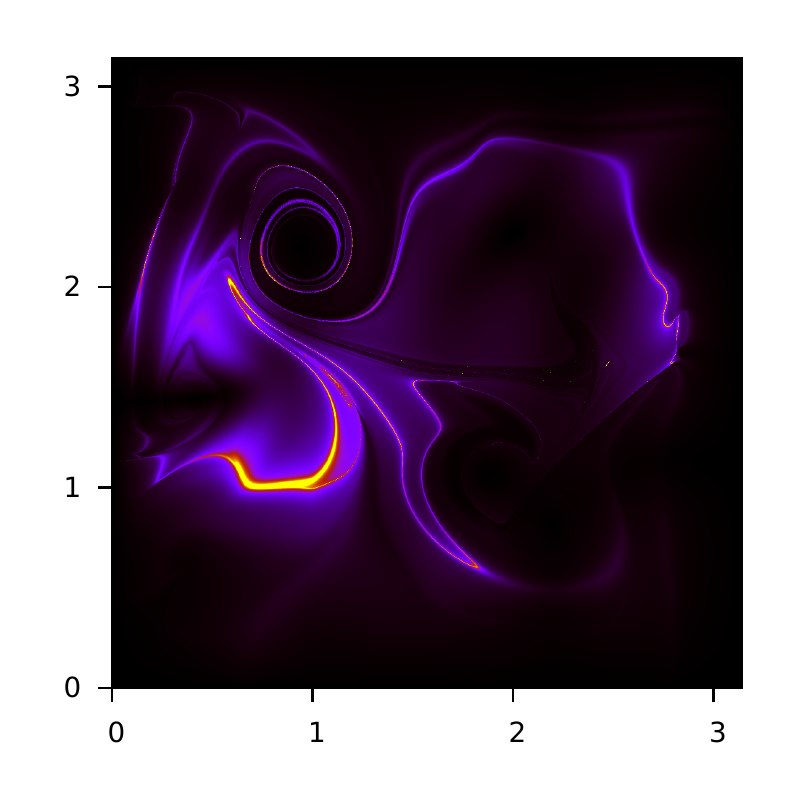}
    \label{fig:map_raw}}
    \subfigure[coarse-grained, \(\varepsilon=32\)]{
    \includegraphics[width=0.3\textwidth]
    {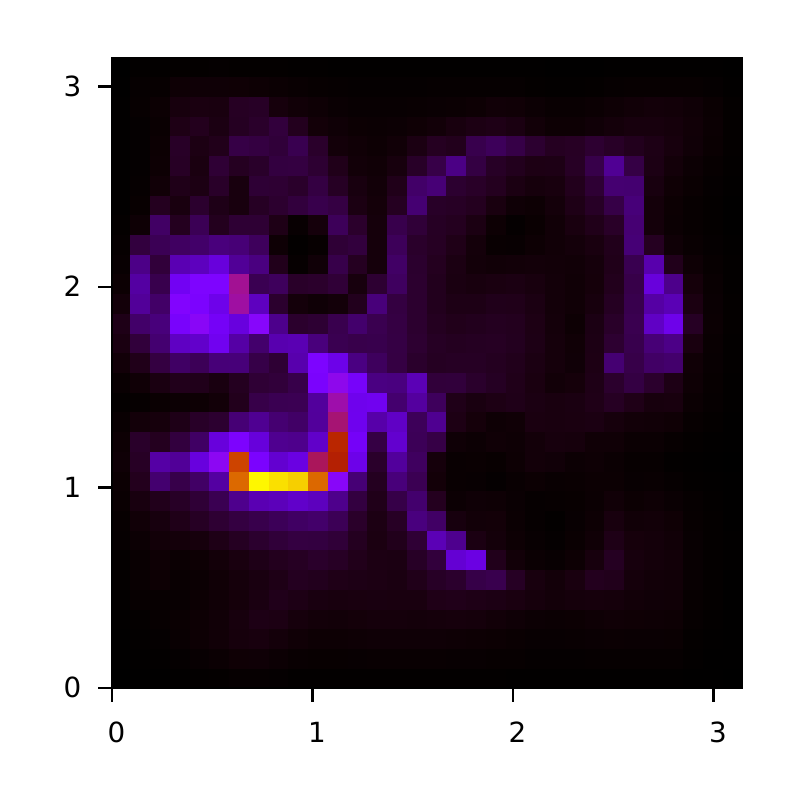}
    \label{fig:map_cg}}
    \subfigure[smoothed]{
    \includegraphics[width=0.3\textwidth]
    {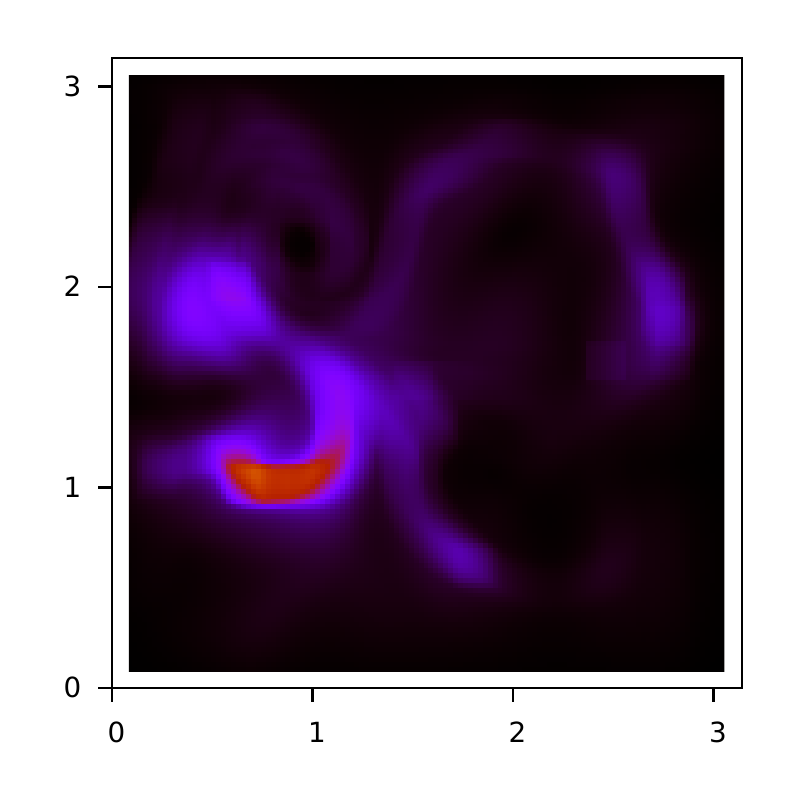}
    \label{fig:map_smoothed}}
    \subfigure[\(\varepsilon=32\)]{
    \includegraphics[width=0.3\textwidth]
    {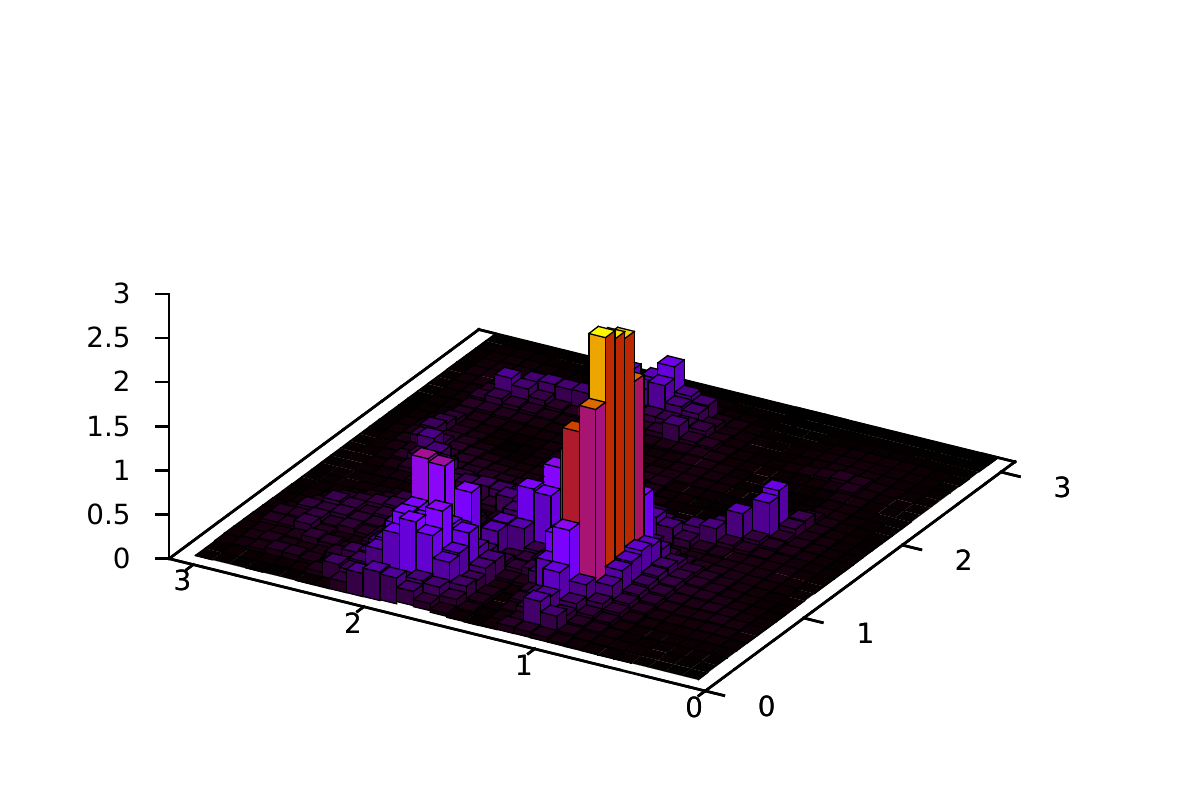}
    \label{fig:cged}}
    \subfigure[single coarse-graining cell]{
    \includegraphics[width=0.3\textwidth]
    {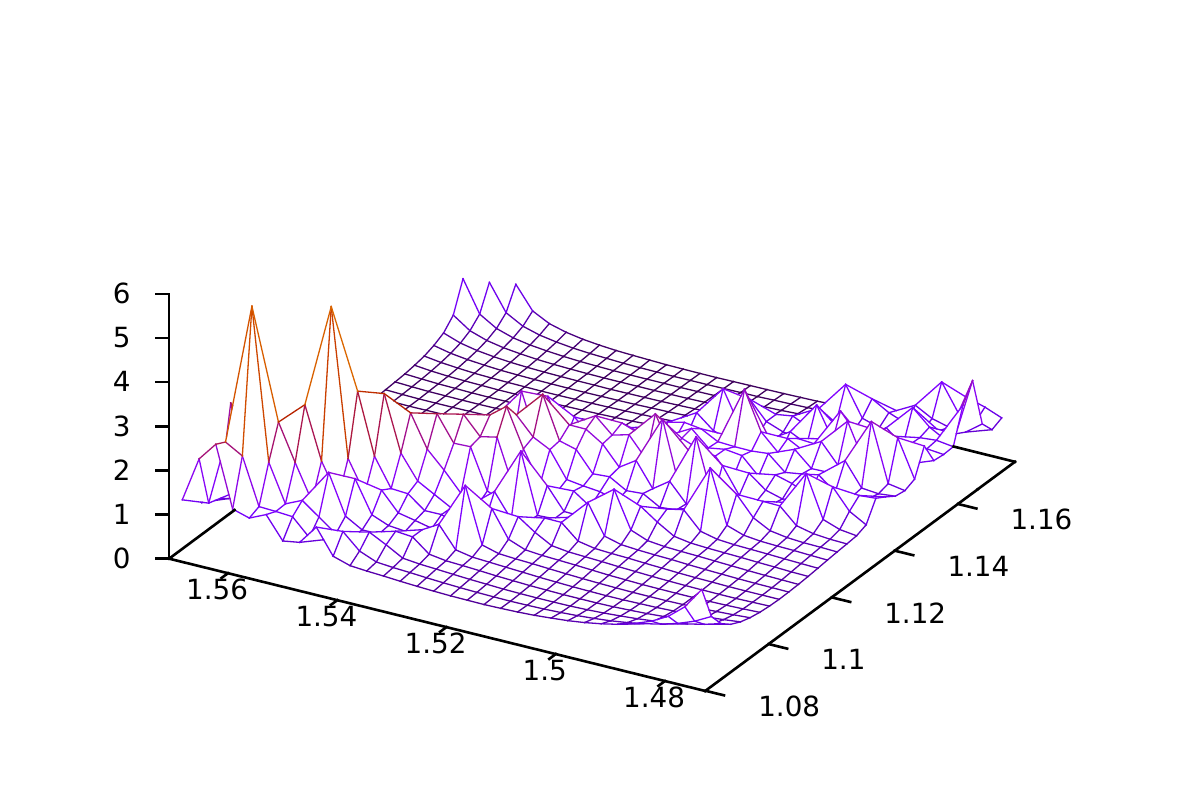}
    \label{fig:raw}}
    \subfigure[smoothing applied]{
    \includegraphics[width=0.3\textwidth]
    {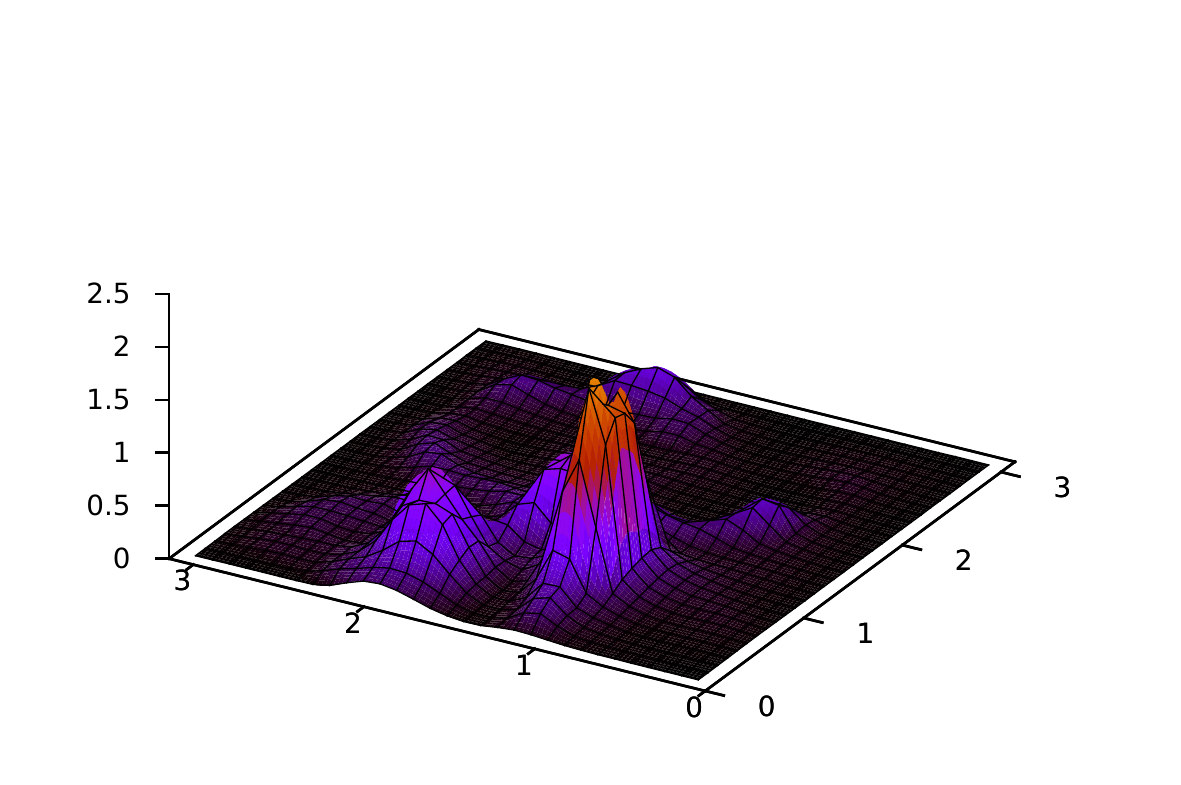}
    \label{fig:smoothed}}\\
    \caption{Figures illustrating the effects of coarse-graining for a 16-mode
    system at \(t = \frac{\pi}{2}\). \ref{fig:map_raw} shows a snapshot of the fine-grained density $\rho$ on a $1024 \times 1024$ lattice; \ref{fig:map_cg} shows the \emph{coarse-grained density} derived from averaging the fine-grained density over square cells containing $32\times 32$ lattice points (in which case we say the system has a `coarse-graining length $\varepsilon = 32$'). \ref{fig:map_smoothed} shows the result of a `smoothed' coarse-graining - using overlapping cells - which is more suitable for plotting graphs~\cite{valentini05}. \ref{fig:cged} is the same coarse-grained density as in Fig.~\ref{fig:map_cg} from a different perspective; \ref{fig:raw} is a close-up
    of a single coarse-graining cell, at the level of individual lattice
    points. The irregular nature of the underlying distribution at this level is
    clear;
    \ref{fig:smoothed} is the same as Fig.~\ref{fig:map_smoothed} from a different perspective.}
    \label{fig:coarsegraining}
  \end{center}
\end{figure*}

The effects of coarse-graining on the particle density at some randomly-selected time may be seen in Fig.~\ref{fig:coarsegraining}.

For the 16-mode case, it was found in Ref.~\onlinecite{valentini05} that \(\cgh\) decays approximately exponentially  as
\begin{eqnarray}
\cgh\left(t\right) \approx \cgh_0 e^{-^{t}/_{\tau}}
\end{eqnarray}
One of us (AV) has presented a theoretical estimate of the relaxation timescale $\tau$ obtained
by considering the behaviour of the second-time derivative of $\cgh$ at $t = 0$
(where $\cgh$ posesses a local maximum)~\cite{valentini_thesis, valentini01}. As
discussed in more detail below, it was shown that in the limit $\varepsilon
\rightarrow 0$, $\tau$ scales inversely with $\varepsilon$. Further estimates,
or simply dimensional analysis, then suggested the rough
formula~\cite{valentini01} 
\begin{equation}
\label{eqn:prediction} \tau \sim \frac{\hbar^{2}}{\varepsilon \,
m^{^1/_2}\left(\Delta E \right)^{^3/_2}}. 
\end{equation}
Here \(\varepsilon\) is the length of the coarse-graining cells and 
\(\Delta E\) is the energy spread of the wave function. For reasonable values of $\varepsilon$, this estimate was in rough agreement with the numerical value~\cite{valentini05}.

However, because $\cgh$ has a local maximum at $t=0$, the estimate on the right-hand side of Eqn.~\ref{eqn:prediction} -- obtained from the second time-derivative of $\cgh$ at $t=0$ -- can only define a timescale that is valid close to $t = 0$. As we shall see, it cannot properly represent the timescale $\tau$ associated with the subsequent (approximately) exponential decay. The results of this paper in fact show a scaling of $\tau$ with $\Delta E$ that disagrees with Eqn.~\ref{eqn:prediction} - but which is in agreement with an improved estimate discussed below.

The question of how $\tau$ varies with the number of modes was not investigated by Valentini and Westman~\cite{valentini05} owing to computational difficulties.
That gap is filled in this paper.

\section{Numerical simulations} 
\label{sec:numerics} 

In this work we compute the dependence of the relaxation time \(\tau\) on the
coarse-graining length \(\varepsilon\) and energy spread \(\Delta E\) through
explicit numerical simulations. An initially non-equilibrium probability density
in a 2D infinite potential square well is evolved according to pilot-wave
dynamics, using a wave function consisting of an out-of-phase superposition of
the first \(M\) energy eigenstates (normal modes). For this choice of wave
function, taking all the modes to have equal weight, we have \(\Delta E \sim
M^{2}\) so we look for a dependence of the form \begin{equation} \label{eqn:tau}
\tau \sim \varepsilon^{q}M^{p} \end{equation} On the basis of
Eqn.~\ref{eqn:prediction}, for example, we would expect \(p = -3\) and \(q=
-1\).

To study this system (and potentially others, since it is designed to be easily
extendible) we have written a new computer code named `LOUIS'~\cite{louis} which
uses pilot-wave dynamics to calculate particle trajectories. Given an initial
probability density and wave function, LOUIS is able to use these trajectories
to compute the probability density function and coarse-grained subquantum
\emph{H}-function at later times. It is a ground-up reimplementation of the code
used in Refs.~\onlinecite{valentini05, colin09} and is up to two orders of
magnitude faster with significantly more capabilities. Currently it can treat
infinite potential square wells in one, two, or three dimensions using a finite
superposition of eigenstates to represent the wave function. The relative
weights and phases of the eigenstates may be specified in the input or chosen
randomly (but reproducibly, using preset seeds). The scale of the
coarse-graining may also be set manually, outputting results for multiple
coarse-graining lengths in a single run of the program.

Since the results of interest involve calculation of the subquantum
\emph{H}-function, which in turn involves a numerical integration over the area
of the potential well, the quantities in the integrand, \(\rho\) and \(\psisq\),
must be evaluated on a regular lattice. In all calculations presented here, a
\(1024 \times 1024\) lattice is used covering a square two-dimensional cell of
length \(\pi\). The `coarse-graining length' \(\varepsilon\) refers to the
number of lattice points along one side of a coarse-graining cell.

\subsection{Details of the algorithm} 

The LOUIS code uses de Broglie-Bohm trajectories to calculate how the particle
probability density evolves from a given initial density. At each of a sequence
of requested times, it evaluates the particle density and wave function at all
points on the fine-grained lattice, and then applies coarse-graining on the
requested scales. The coarse-grained \emph{H}-function is calculated from these
data at each timestep, and output files containing \(\left( t,H \right)\) pairs
and \(\left( t,\ln H \right)\) pairs are written to disk. The program calculates
a straight-line fit using linear regression of the \(t\) vs.\ \(\ln H\) data;
assuming exponential scaling, the gradient of this is the decay constant or
relaxation time \(\tau\).  

How do we calculate the density at a later time? We have seen that the ratio
\(\frac{\rho}{\psisq}\) is preserved along trajectories, implying that the
density at position \(\bx\) and time \(t\) may be calculated
from 
\begin{equation} 
\rho \! \left( \bx, t \right) = \frac{ \left| \Psi \!
\left( \bx, t \right) \right|^{2} }{ \left| \Psi \! \left( \bx_{0}, 0
\right) \right|^{2} }\rho \! \left( \bx_{0}, 0 \right) 
\end{equation} 
where the positions \(\bx_{0}\) and \(\bx\) are points on the same trajectory,
at times \(0\) and \(t\) respectively. The value of \(\psisq\) can be calculated
analytically at all positions and times, and \(\rho\!\left({\bf x}_0, 0\right)\)
is a known function, therefore $\rho({\bf x},t)$ may be calculated directly once
we know the trajectory endpoint ${\bf x}$. This is the crucial relation used to
calculate probability density functions from trajectories.

In fact, certain practicalities require real calculations to be performed in a
slightly different manner. The subquantum \emph{H}-function is evaluated through
numerical integration over the 2D box from a set of values of $\rho$ and
$|\Psi|^2$ calculated at discrete points. Since accurate and efficient
quadrature algorithms in few dimensions generally require the points to be
sampled uniformly across the region, LOUIS starts with a uniform lattice at time
\(t\) and exploits the time-reversibility of the dynamics to calculate particle
trajectories \emph{backwards in time} to $t=0$. This ensures uniform sampling of
$\rho({\bf x},t)$ at $t$ when the quadrature is to be performed. This has the
unfortunate consequence that if \(\rho\) is required at a later time
\(t^\prime>t\) this `backtracking' has to be done all the way to $t=0$ again:
the data calculated at time \(t\) cannot be used again.

The rate-limiting step of the LOUIS program is the numerical integration of the
de Broglie guidance equation ${\bf v({\bf x}, t)} = \nabla S({\bf x},t)$ (in
atomic units) to compute the particle trajectories ${\bf x}(t)$. One may use a
variety of standard algorithms; an excellent choice for these purposes is
Runge-Kutta-Fehlberg~\cite{numerical_recipes}. Currently, the Schr\"odinger
equation is not integrated numerically to compute the time-development of the
wave function; instead, only finite superpositions of stationary states are used
so the wave function can be evaluated exactly for any $t$.

The velocity of the particle at any point may be computed from $\mathrm{Im}\frac{\nabla \Psi({\bf x},t)} {\Psi({\bf x},t)}$ where the $M$-mode wave function is given by
\begin{eqnarray}
\label{eqn:wavefunc}
\Psi({\bf x},t) = \frac{2}{\pi \sqrt{M}}\sum_{m,n=1}^{\sqrt{M}} \sin(mx)\sin(ny) \exp{i(\theta_{mn} - E_{mn}t)}.
\end{eqnarray}
Here $E_{mn}$ are the energy eigenvalues $\frac{1}{2}(m^2+n^2)$, the $\theta_{mn}$ are the (randomly chosen) initial phases, $m,n = 1,2,\cdots,\sqrt{M}$ are positive integers, and (for convenience) $M$ has an integer square root.

As with all such algorithms, Runge-Kutta-Fehlberg basically involves adding
small increments to a function - here ${\bf x}(t)$ - where the increments are
given by derivatives ($\frac{\mathrm{d}{\bf x}}{\mathrm{d}t} = {\bf v}= \nabla S
= \mathrm{Im}\frac{\nabla \Psi}{\Psi}$) multiplied by variable step sizes (here,
a timestep $\Delta t$). In order to increase the accuracy, a tolerance is set
for the maximum error on each step (the \emph{step tolerance}); if the error is
greater than this, a smaller timestep is used (subject to appropriate underflow
checks). When the integration has been performed along the entire trajectory
between the required initial and final times, the whole trajectory is recomputed
with the step tolerance decreased by a factor of \(10\). If the two final
positions agree within a certain tolerance (the \emph{trajectory tolerance}),
then the trajectory is kept. If not, the process is repeated with smaller and
smaller step sizes until the trajectories converge, or until the step tolerance
reaches a certain minimum value where the calculation will take too much time
and the trajectory is flagged as \emph{failed}. Failed trajectories are not used
in the subsequent computation of the density. The proportion of failed trajectories rarely exceeds \(1\) in
\(1000\) and their contribution to the overall error is negligible. In general,
they are trajectories that come too close to a wave function node (where the
velocity field diverges).

Computational cost is the main limiting factor. The calculation of relaxation
timescales is very computationally intensive, requiring many long,
high-precision numerical integrations. Since the particle wave function and its
gradient must be evaluated at each step in the integration of each trajectory,
the complexity of the wave function contributes significantly to the time taken
to perform the calculation; runs with larger numbers of modes in the
superposition are considerably more expensive. For example, a typical
calculation on an elderly cluster of sixteen processors (seventeen evaluations
of \(\cgh\) between time \(0\) and time \(4\pi\)) took \(9.6\) CPU-hours with a
9-mode wave function. A comparable calculation with 36 modes took \(542\)
CPU-hours. The calculations we report here are for 4, 9, 16, 25, 36, 49, and 64
modes. Previous calculations were done exclusively with either 4 modes (Colin
and Struyve~\cite{colin09}) or 16 modes (Valentini and
Westman~\cite{valentini05}). 

\section{Results and discussion}
\label{sec:results} 
\begin{figure*}[p]
  \begin{center}
    \subfigure[varying coarse-graining length, \(\varepsilon\)]{
    \includegraphics[width=0.3\textwidth]
    {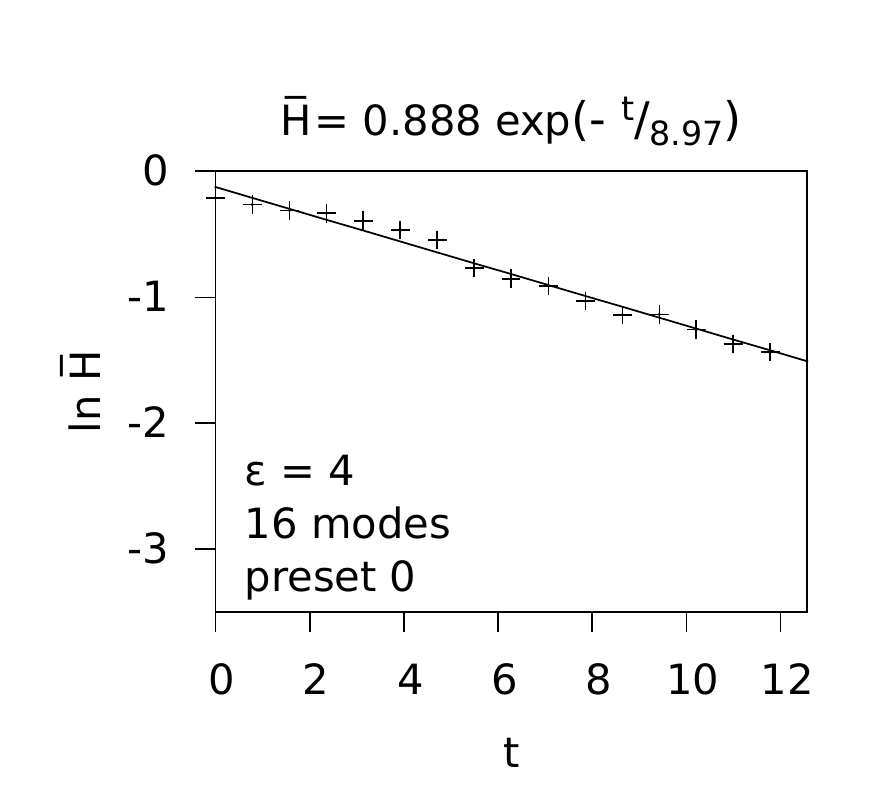}
    \includegraphics[width=0.3\textwidth]
    {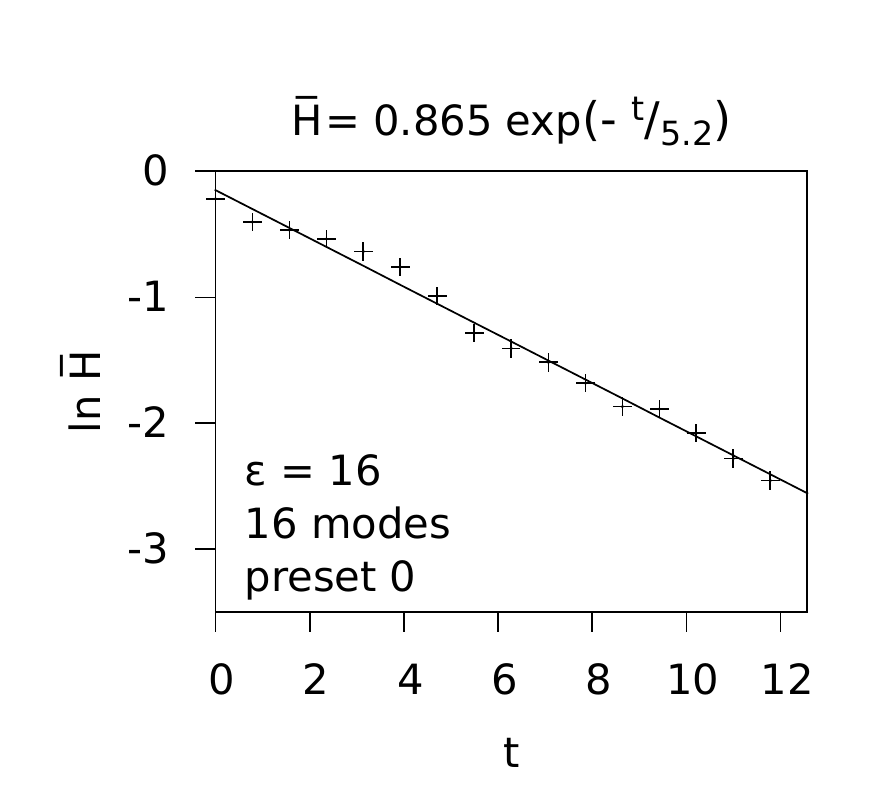}
    \includegraphics[width=0.3\textwidth]
    {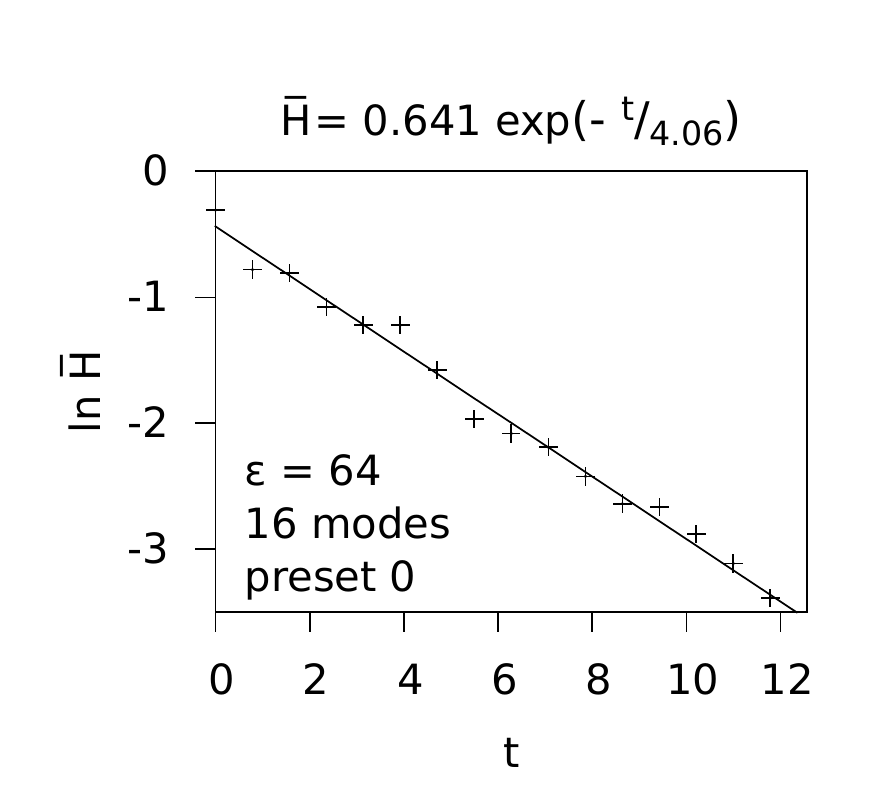}}\\
    \rule{0cm}{1.0cm}\\
    \subfigure[varying number of modes, \(N\)]{
    \label{fig:nmodes}
    \includegraphics[width=0.3\textwidth]
    {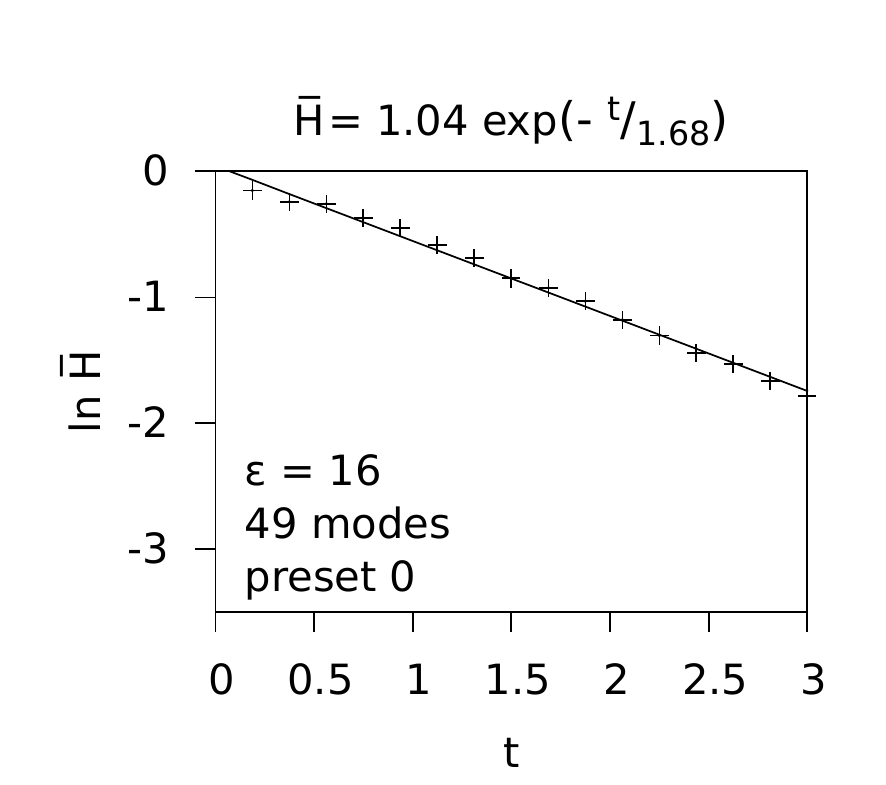}
    \includegraphics[width=0.3\textwidth]
    {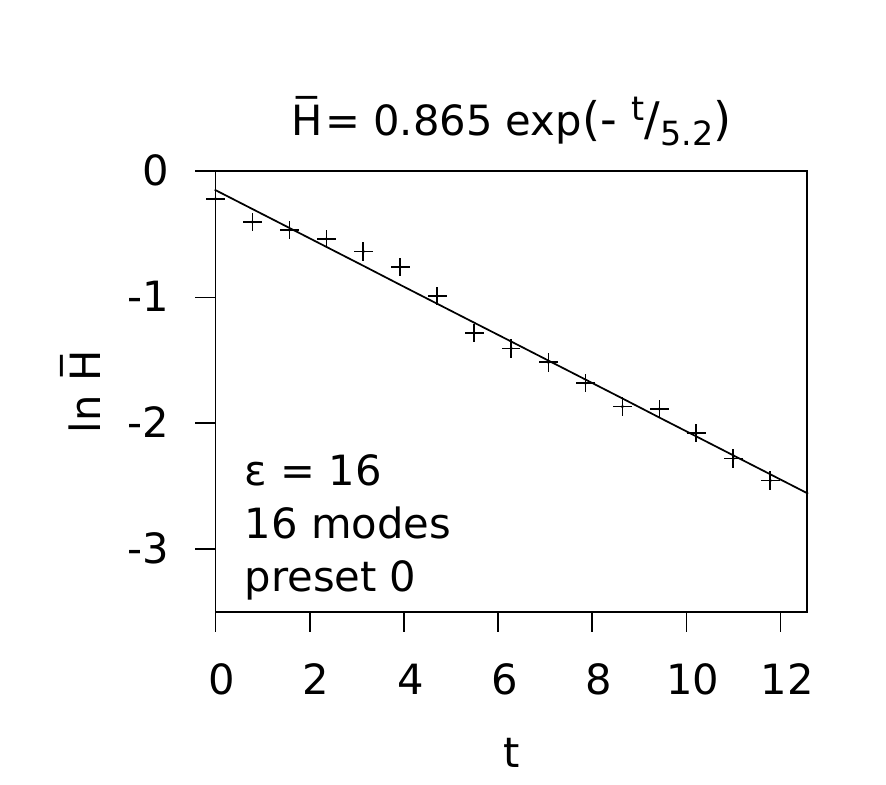}
    \includegraphics[width=0.3\textwidth]
    {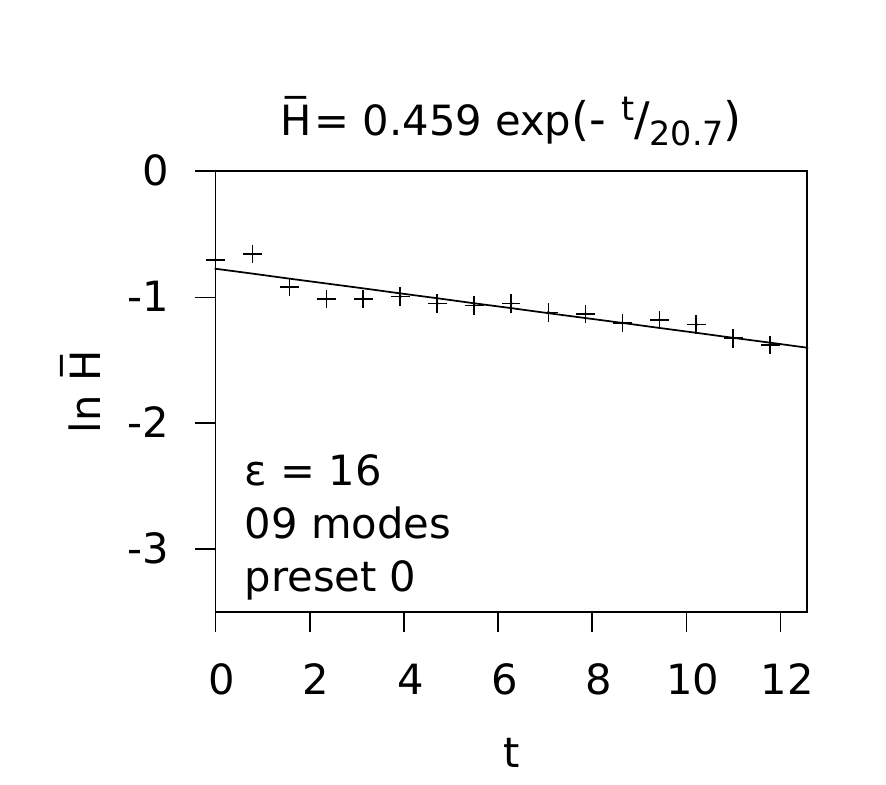}}\\
    \rule{0cm}{1.0cm}\\
    \subfigure[varying initial phases]{
    \includegraphics[width=0.3\textwidth]
    {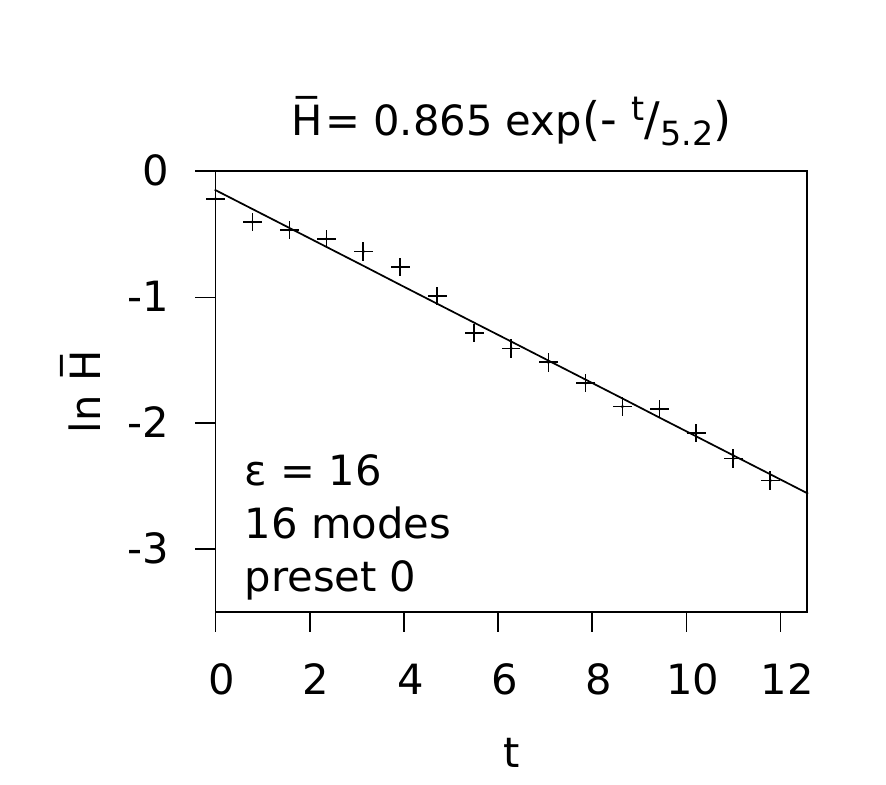}
    \includegraphics[width=0.3\textwidth]
    {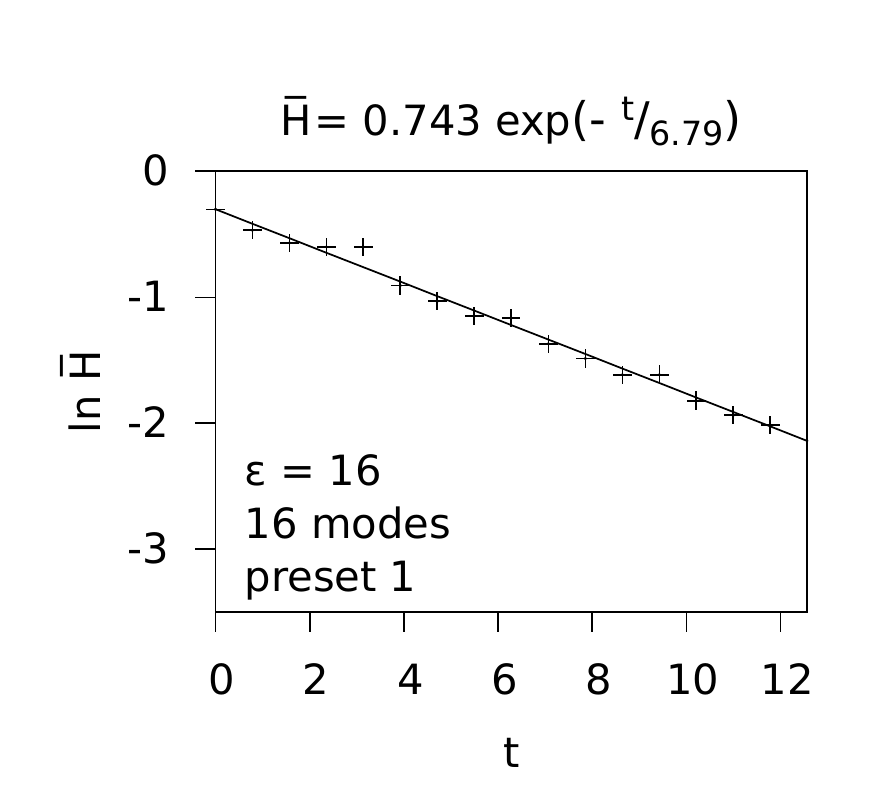}
    \includegraphics[width=0.3\textwidth]
    {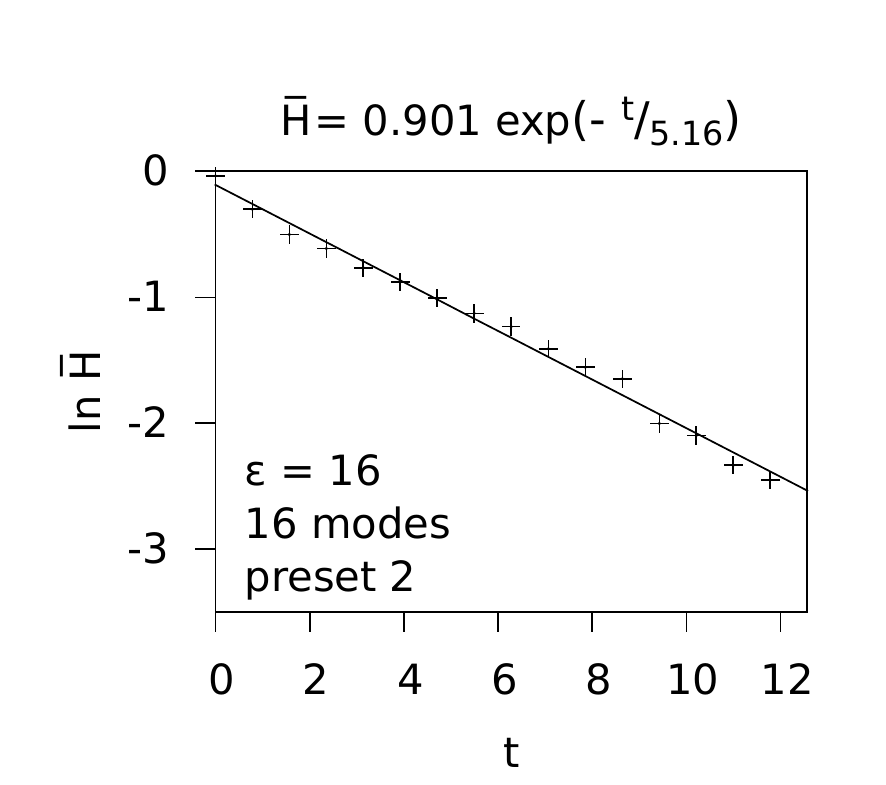}}\\
    \rule{0cm}{1.0cm}\\
    \caption{Graphs showing that the model of exponential decay is robust
    under the conditions studied in this article. The horizontal axis on the
    first graph in \ref{fig:nmodes} is scaled differently to the others, to
    better fit the range of data.}
    \rule{0cm}{2cm}\\
    \label{fig:exponential}
  \end{center}
\end{figure*}

We begin by verifying that exponential decay is an appropriate model for the
evolution of \(\cgh\) under these conditions as was assumed in the definition of
the relaxation timescale.  Fig.~\ref{fig:exponential} shows plots of $\ln \cgh$
vs. $t$ for different coarse-graining lengths \(\varepsilon\), different numbers
of modes \(M\) in the superposition,  and different sets of initial phases
$\theta_{mn}$ in the wave function (in order to obtain reproducible results the
phases are fixed by a single `preset' parameter in the LOUIS input file, which
controls the seed for the random generation of a set of phases). In all cases we
find a good straight-line fit to the data, validating the assumption of
exponential decay in \(\cgh\) over this range of conditions. This was not
unexpected as exponential decay was previously demonstrated by Valentini and
Westman~\cite{valentini05}, though only for the case of \(M = 16\) modes with a
fixed coarse-graining length.   

\subsection{Relaxation time $\tau$ as a function of the number of modes $M$} 

\begin{figure*}[p]
  \begin{center}
    \includegraphics[width=0.4\textwidth]
    {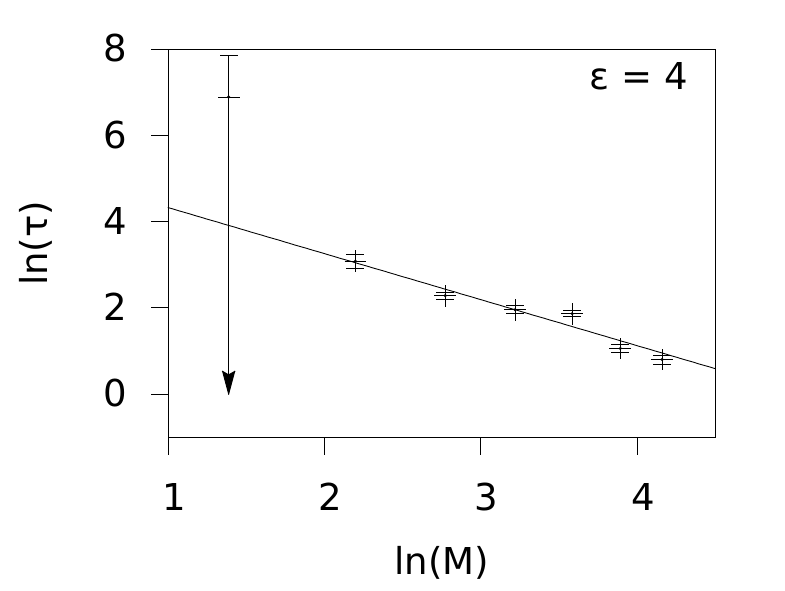}\hspace{0.1\textwidth}
    \includegraphics[width=0.4\textwidth]
    {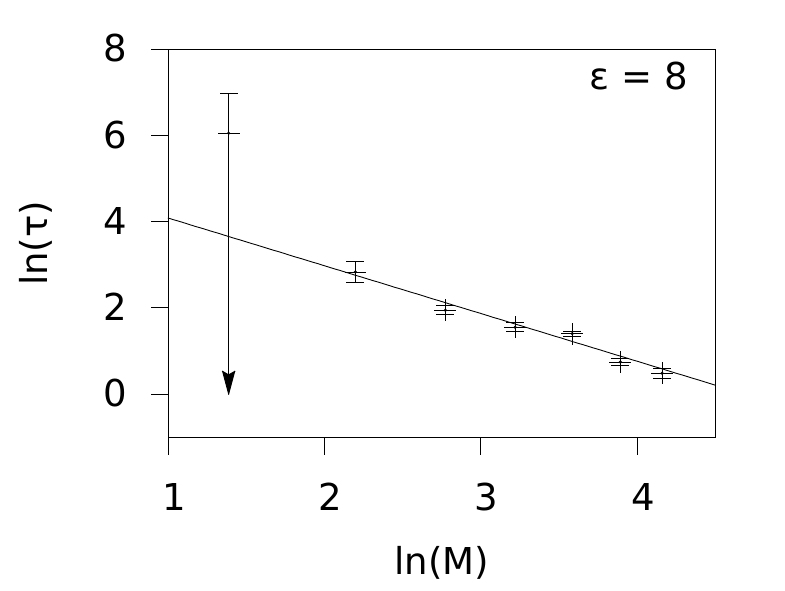}\\
    \includegraphics[width=0.4\textwidth]
    {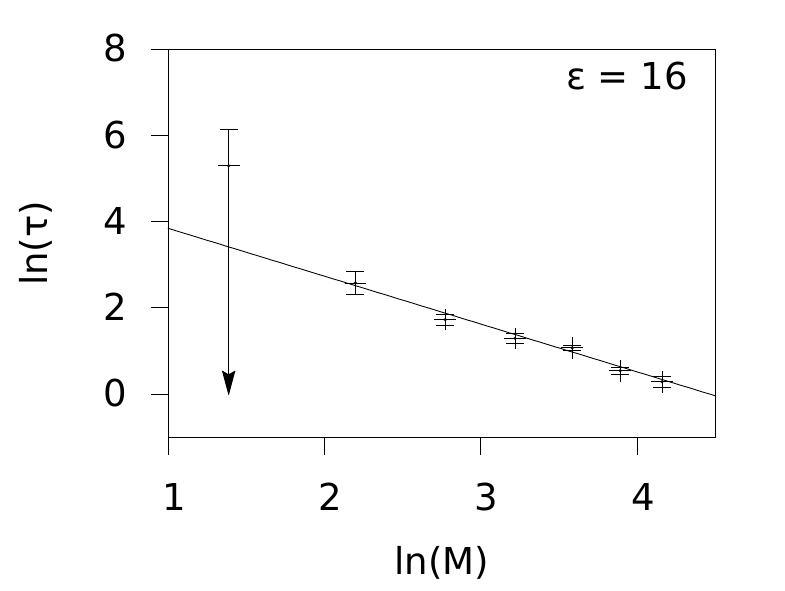}\hspace{0.1\textwidth}
    \includegraphics[width=0.4\textwidth]
    {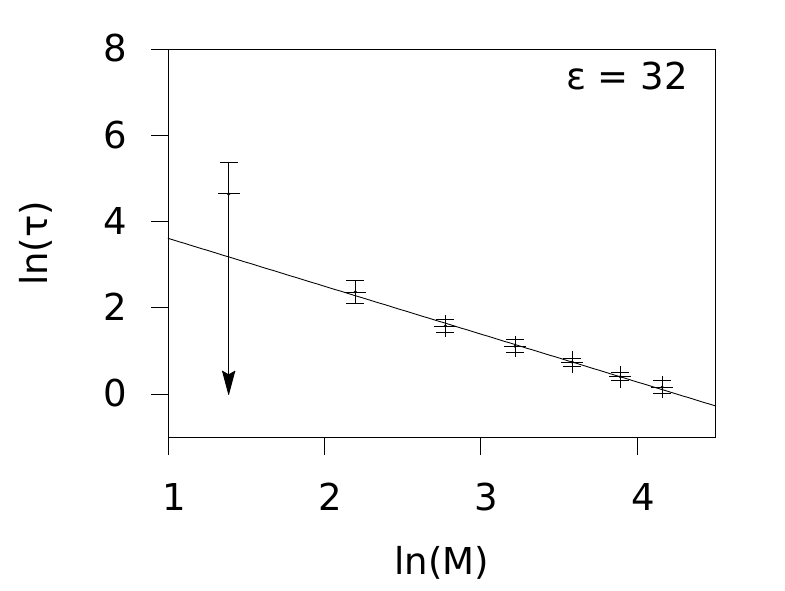}\\
    \includegraphics[width=0.4\textwidth]
    {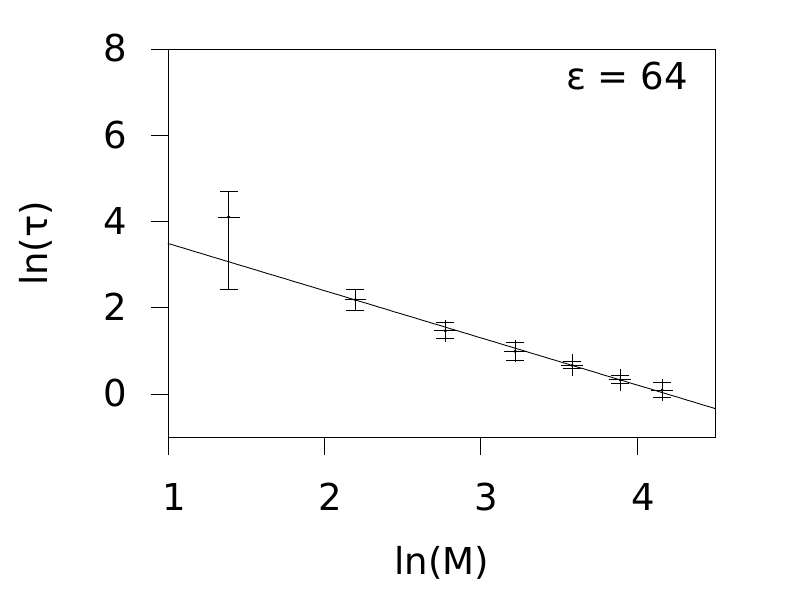}
    \rule{0.5\textwidth}{0.0pt}\\
    \caption{Graphs showing \(\ln \tau\) against \(\ln M\), for varying coarse-graining lengths.
    Logarithmic axes have been used to identify a power law
    relationship, \(\tau \propto M^{p}\). The error bars were estimated by
    running LOUIS six times for each \(\left( \varepsilon,M \right)\) pair, using
    different initial phases. The values shown above correspond to the mean of
    these six runs, and the error bars are one standard deviation.
    The errors on the points for \(M
    = 4\) cannot be properly represented on a logarithmic scale, as the
    lower bound is less than zero. There is reason to believe (see
    section~\ref{sec:results}) that these points may be excluded, and with such
    a large error their weight in fitting would be very small, so
    Fig.~\ref{fig:tau_vs_n} shows the same results without these points.}
    \label{fig:tau_vs_n_raw}
  \end{center}
\end{figure*}

Fig. \ref{fig:tau_vs_n_raw} shows plots of  \(\ln \tau\) against \(\ln M\) for
various coarse-graining lengths \(\varepsilon\) where now logarithmic axes
are used in order to search for a power law relationship of the form  $\tau
\propto M^{p}$. Error bars were calculated by running LOUIS six times for each
\(\left( \varepsilon,M \right)\) pair with different initial phases
$\theta_{mn}$ in each run; the mean of these was taken to be the best estimate
for the timescale with the standard deviation taken to be the error bar. 

In the case of 4-mode simulations,  the spread of values of \(\tau\) was so
large that the error bar \(\Delta \tau\) could not be displayed on a logarithmic
plot (hence the arrow at the base of the corresponding error bars in
Fig.~\ref{fig:tau_vs_n_raw}).  Some representative values of \(\tau\) were \(
\left( 980 \pm 1600 \right) \) for \(\varepsilon = 4\) and \( \left( 100 \pm 110
\right) \) for \(\varepsilon = 32\).  This large spread in the timescale for low
\(M\) can be understood by considering the role of wave function nodes in the
relaxation. The rapidly-varying velocity field in the vicinity of nodes is
believed to be a significant driving force for this process~\cite{valentini05},
and so the initial positions of the nodes are likely to affect the timescale
(these initial positions are moved around when modifying the set of initial
phases $\theta_{mn}$). The important point is that in larger superpositions with
larger \(M\) there are many nodes and the exact change in positions of nodes
will have less effect because the average distribution will be similar. With a
small superposition, maybe only containing one node, the initial position of
this node will have a much larger effect on the subsequent relaxation, and so
the six runs with different initial phases will tend to produce very different
results. 

\begin{figure*}[p]
  \begin{center}
    \subfigure[]{
    \label{fig:cg04}
    \includegraphics[width=0.4\textwidth]
    {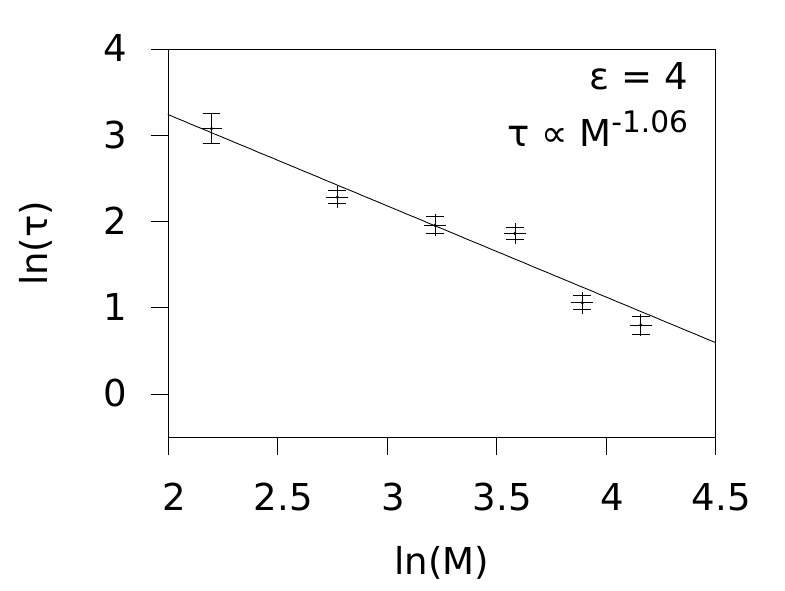}}\hspace{0.1\textwidth}
    \subfigure[]{
    \label{fig:cg08}
    \includegraphics[width=0.4\textwidth]
    {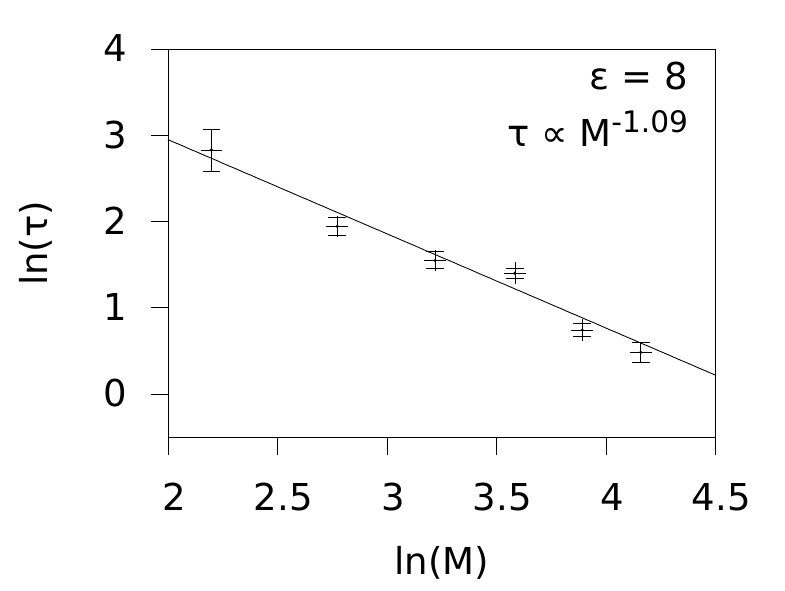}}\\
    \subfigure[]{
    \label{fig:cg16}
    \includegraphics[width=0.4\textwidth]
    {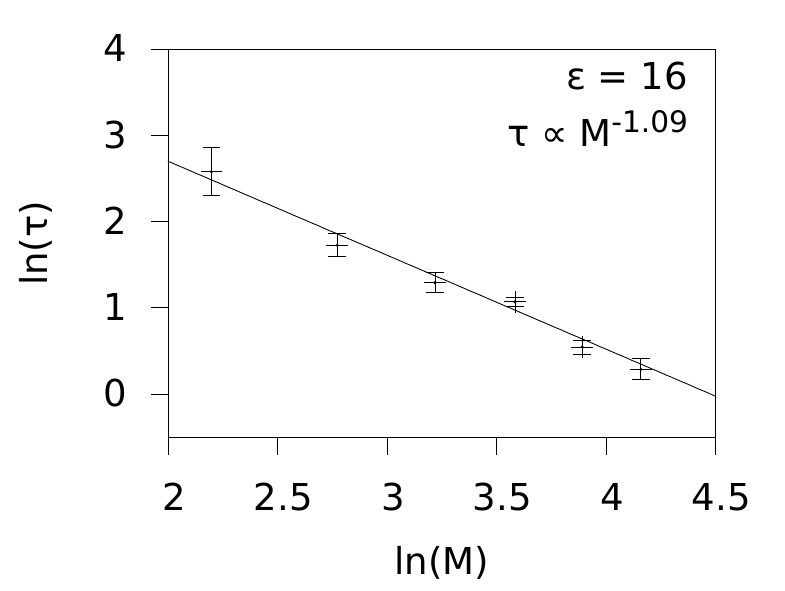}}\hspace{0.1\textwidth}
    \subfigure[]{
    \label{fig:cg32}
    \includegraphics[width=0.4\textwidth]
    {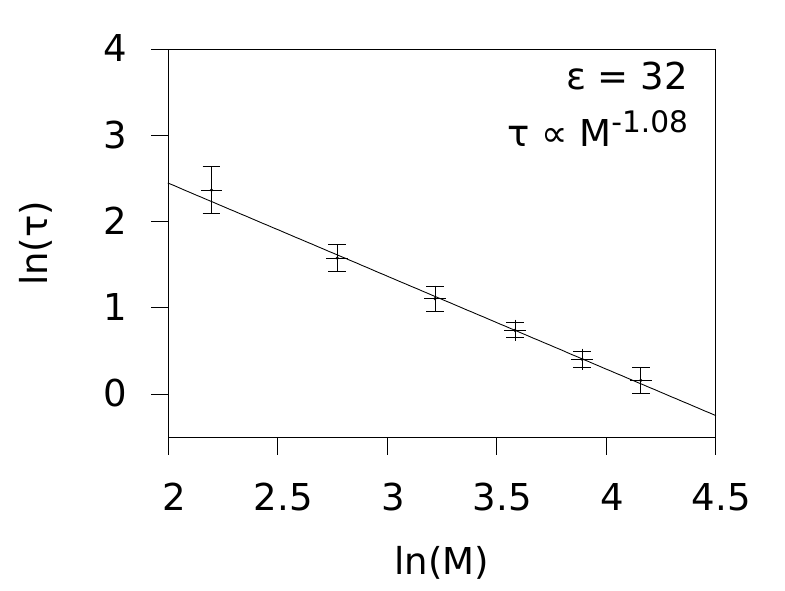}}\\
    \subfigure[]{
    \label{fig:cg64}
    \includegraphics[width=0.4\textwidth]
    {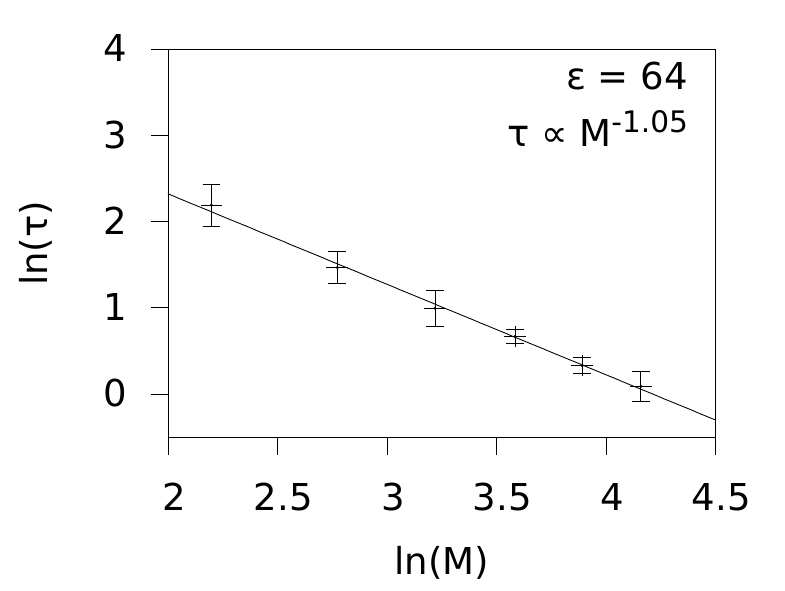}}\hspace{0.1\textwidth}
    \subfigure[]{
    \label{fig:alle}
    \includegraphics[width=0.4\textwidth]
    {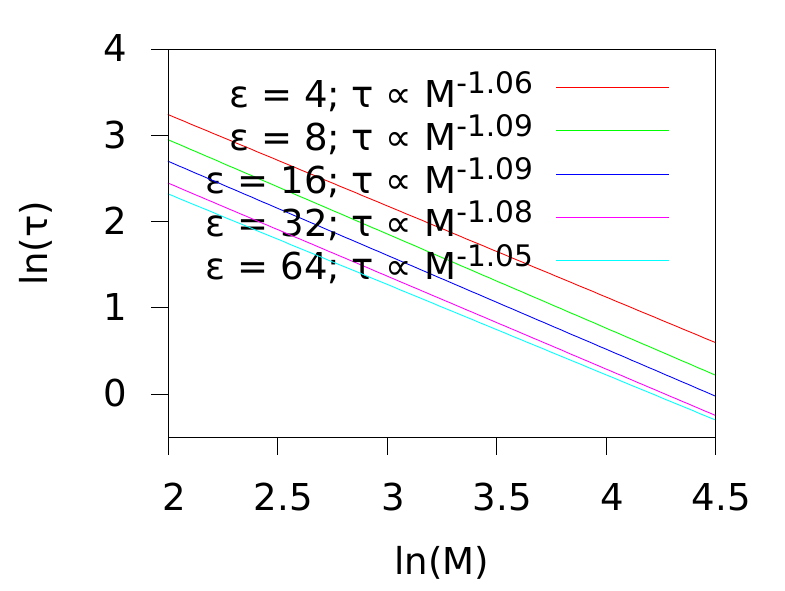}}\\
    \caption{Graphs of \(\tau\) against \(M\) for all \(\varepsilon\), excluding
    the points for \( M = 4 \). The equations for the best power law fit (using
    the GNUplot implementation of the Marquardt-Levenberg nonlinear
    least-squares fitting algorithm) is shown on each of the graphs.
    \ref{fig:alle} shows a comparison of the best
    fits for all \(\varepsilon\). It can be seen that for all \(\varepsilon\)
    there
    is a roughly constant power law with index \(\approx -1\).}
    \label{fig:tau_vs_n}
  \end{center}
\end{figure*}

A full animation of the relaxation process can be seen online at Ref.~\cite{density_video}. 
The swirling
vortices in the density surrounding the moving wave function nodes are quite
striking, and one can obtain more of a visual sense of why the presence of
nodes increases the chaotic nature of the trajectories.

In Fig.~\ref{fig:tau_vs_n}, therefore, we choose to plot the same data as in 
Fig.~\ref{fig:tau_vs_n_raw} but with the points for \(M = 4\) omitted, in an
attempt to show the relationship between \(\tau\) and \(M\) in the regime where
the distribution of nodes may be considered roughly fixed.  Can we justify this?
Arguably, points with such a large error will have little effect on the curve
fitting, as points are weighted according to the inverse of their error.
However, a better reason for neglecting the points for \(M=4\) is that the
theoretical predictions are actually in terms of \(\Delta E\), the uncertainty
in the energy of the wave function, rather than the number of modes, \(M\), in
the wave function. The approximation used in this analysis is that, \(\Delta E
\sim M^{2}\), which only applies for larger \(M\).

The best-fitting power law is shown on the graph in each case, and they are also
summarised in Table~\ref{tab:tau_vs_n}. Clearly these results do \emph{not}
support the theoretical prediction (described as a `crude estimate' in
Ref.~\cite{valentini05}) that the relaxation time should be  proportional to
$M^{-3}$. Instead they very strongly suggest a relationship of the form \(\tau
\propto M^{-1}\), to within the estimated error. This suggests that some of the
approximations made in obtaining the prediction in Eqn.~\ref{eqn:prediction}
are invalid. 

The relevant arguments used to obtain the apparently incorrect estimate of the scaling are set out in Ref.~\cite{valentini01}. They
begin by defining a relaxation timescale $\tau$ in terms of the rate of decrease of
$\cgh$ near $t=0$ via the following formula:
\begin{equation}
\label{eqn:one_over_tausquared}
  \frac{1}{\tau^{2}} \equiv - \frac{1}{\cgh_0} \left( \frac{\mathrm{d}^{2}
  \cgh}{\mathrm{d}t^{2}} \right)_{0}.
\end{equation}
The second derivative is used rather than the first derivative since \(\left(
\frac{\mathrm{d}\cgh}{\mathrm{d}t} \right)_{0} = 0\) (as may be shown
analytically~\cite{valentini_thesis}. The $\cgh$-curve necessarily has a local maximum at
$t=0$. This property might, at first sight, seem to be incompatible with the
observed exponential decay. But in fact, it must be the case that the
exponential decay sets in soon after $t=0$, and that in the limit $t \rightarrow
0$ the decay is not exponential. The timescale of
Eqn.~\ref{eqn:one_over_tausquared} then applies only to this very short period
immediately after $t=0$. In other words, while
Eqn.~\ref{eqn:one_over_tausquared} may well be a good estimate for the
relaxation timescale in the limit $t \rightarrow 0$, it cannot accurately
estimate the time constant in the exponential tail -- where the bulk of the
relaxation takes place -- and is therefore of little practical relevance.
  \begin{SCtable*}[2.5][t]
    \begin{tabular}{|r|c|}
      \hline
      \hspace{.5cm}\(\phantom{M}\varepsilon\)\hspace{.5cm} & \(p\)\\
      \hline
      4 & \(-1.06 \pm 0.18\)\\
      \hline
      8 & \(-1.09 \pm 0.17\)\\
      \hline
      16 & \(-1.09 \pm 0.12\)\\
      \hline
      32 & \(-1.08 \pm 0.03\)\\
      \hline
      64 & \(-1.05 \pm 0.03\)\\
      \hline
    \end{tabular}
    \caption{Summary of results for a relationship between \(\tau\) and
    \(M\), for various coarse-graining lengths. The error is estimated from
    the straight line fit, which takes into account the errors on the points
    used for the fit. The values are clearly not compatible with \(\tau \propto
    M^{-3}\) as
    in Eqn.~\ref{eqn:prediction}, but are compatible with \(\tau
    \propto M^{-1}\), within error, supporting the relation in
    Eqn.~\ref{eqn:explanation}.}
    \label{tab:tau_vs_n}
  \end{SCtable*}
   
\begin{figure*}[t]
  \begin{center}
    \subfigure[fine-grained]{
    \includegraphics[width=0.4\textwidth]
    {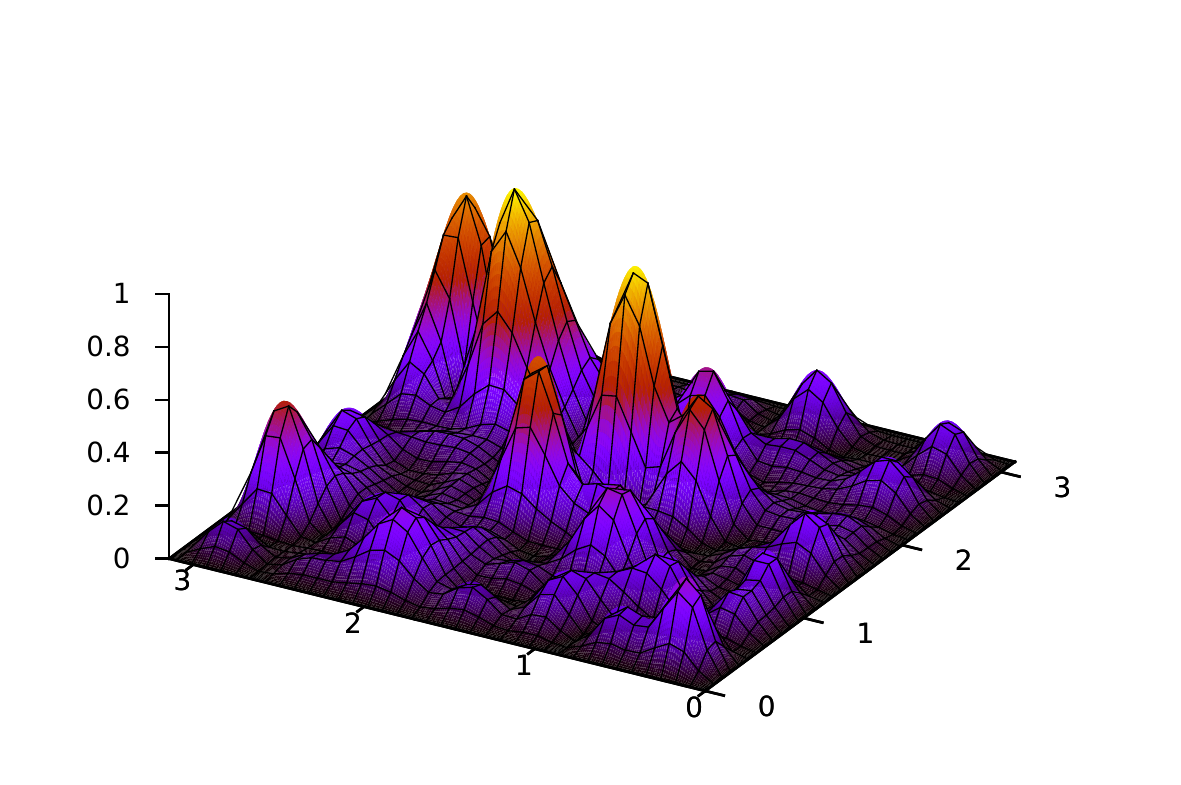}}
    \subfigure[coarse-grained]{
    \includegraphics[width=0.4\textwidth]
    {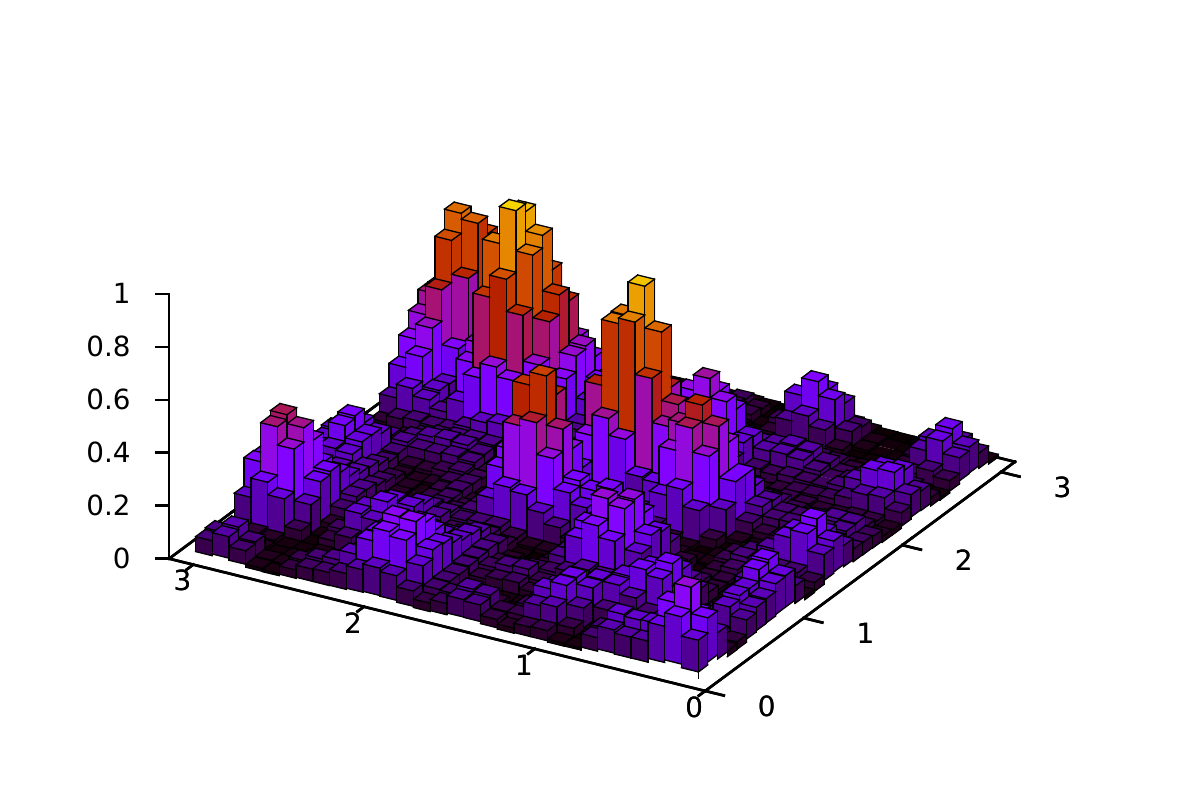}}
    \caption{Comparison of the fine-grained \(\psisq\)
    and the coarse-grained approximation,
    demonstrating that there may be significant variation over the
    length of coarse-graining cells.}
    \label{fig:largecg}
  \end{center}
\end{figure*}

Another potential conflict in the derivation of Eq.~\ref{eqn:prediction} is the
requirement for a coarse-graining length $\varepsilon$ so small that the
velocity field varies little over the length of a coarse-graining cell. The
derivation works by considering the dependence of \(\tau\) on \(\varepsilon\) in
the limit where \(\varepsilon \rightarrow 0\), then applying dimensional
analysis to find the other dependencies. The inverse dependence of $\tau$ (as
defined by Eqn.~\ref{eqn:one_over_tausquared}) on $\varepsilon$ must hold in this limit,
since we have an analytic proof of it. However, unfortunately, the limit seems
too restrictive to be of practical use. In the cases studied here the wave
function apparently varies rapidly enough on that scale - particularly with
larger numbers of modes in the superposition - to ensure that  we are not
working in this limit at all. For example, Fig.~\ref{fig:largecg} demonstrates
the effect of coarse-graining (\(\varepsilon = 32\)) on one of our 64-mode wave
functions. This shows the magnitude of \(\Psi\), rather than the velocity field,
but the length scale over which the latter varies should be even smaller than
the length scale over which the former varies. We may thus conclude that the
velocity field is likely to vary significantly over the length of one
coarse-graining cell, at least under some of the conditions studied here. Taking
all these facts into consideration, it is not surprising that the results do not
confirm Eqn.~\ref{eqn:prediction}, whose domain of validity is probably simply
too narrow to be of practical use.

We now provide a theoretical justification for the observed relation \(\tau \propto M^{-1}\). Let $q_r (r=1,2)$ denote $x$ or $y$. We shall proceed by considering an upper bound on the (equilibrium) mean displacement \(\delta
q_{r}\) of
particles over an arbitrary time interval \(\left[ t_i, t_f \right]\). A relaxation time
may then be defined, in the case of the infinite potential square well, by the condition
that relaxation will occur over timescales $\tau$ such that
the said upper bound becomes of the order of the width $L$ of the potential well. As we shall now show, the timescale will then be
\label{sec:explanation}
\begin{equation}
  \label{eqn:explanation}
  \tau \sim L \sqrt{ \frac{m}{2 \bar{E}}}
\end{equation}
where \(\bar{E}\) is the mean energy, and \(\bar{E} \propto M^{2}\).

There now follows a derivation of the upper bound on the mean displacement \(\delta q_{r}\). The derivation is based on Ref.~\onlinecite{valentini08}, but with some differences that are highlighted below.

First, note that the final displacement \(\delta q_{r} \left( t_f
\right)\) has modulus
\begin{equation}
  \left| \delta q_{r} \left( t_f \right) \right| \leq \int_{t_i}^{t_f} \dt
  \left| \dot q_{r} \left(t\right) \right|
\end{equation}
where \(\dot{q}_{r}\) is the component of the de Broglie-Bohm velocity in the
\(q_{r}\) direction.

The equilibrium mean \(\langle \left| \delta q_{r} \left( t_f \right) \right|
\rangle_{eq}\) then satisfies
\begin{equation}
  \lang \left| \delta q_{r} \left( t_f \right) \right| \rang_{eq} \leq
  \lang \int_{t_i}^{t_f} \dt \left| \dot{q}_{r} \left( t \right) \right|
  \rang_{eq} =
  \int_{t_i}^{t_f} \dt \lang \left| \dot{q}_{r} \left( t \right)
  \right| \rang_{eq}.
\end{equation}
The equilibrium mean speed, \(\lang \left| \dot{q}_{r} \left( t \right)
\right| \rang_{eq}\) is
\begin{equation}
  \lang \left| \dot{q}_{r} \left( t \right) \right| \rang_{eq} = \int \int
  \dx\dy \left| \Psi \left(x,y,t\right) \right|^{2} \left| \dot{q}_{r}
  \left(x,y,t\right) \right|.
\end{equation}
Using the fact that, for any \(x\), \(\lang x \rang \leq \sqrt{ \lang x^{2}
\rang}\), we have
\begin{equation}
  \lang \left| \delta q_{r} \left( t_{f} \right) \right| \rang_{eq} \leq
  \int_{t_i}^{t_f} \dt \sqrt{ \lang \left| \dot{q}_{r} \left(t\right)
  \right|^{2} \rang_{eq}}.
\end{equation}
Now note that, starting from the guidance equation:
\begin{equation}
 \begin{split}
    m^{2} \lang \left| \dot{q}_{r} \right|^{2} \rang_{eq}
    &= \lang \left( \frac{ \partial S }{ \partial q_{r} } \right)^{2}
    \rang_{eq}\\
    &= \int \int \dx \dy \left| \Psi \left( x,y,t \right) \right|^{2} \left(
    \frac{ \partial S \left( x,y,t \right)}{ \partial q_{r} } \right)^{2}\\
    &= \lang \hat{\pi}_{r}^{2} \rang - \int \int \dx \dy \left( \frac{
    \partial \left| \Psi \left( x,y,t \right) \right| }{ \partial q_{r}
    }\right)^{2}
  \end{split}
 \end{equation}
where \(\hat{\pi}_{r}\) is the momentum operator conjugate to
\(q_{r}\), and \(\lang \hat{\pi}_{r} \rang\) denotes the usual quantum
expectation value for the operator \(\hat{\pi}_{r}\). The last equality
follows from the relations
\begin{align}
  \begin{split}
    \lang \hat{\pi}_{r}^2 \rang &= \int \int \dx \dy \Psi^{*} \left( -\frac{
    \partial^{2} }{ \partial q_{r}^{2} } \right) \Psi\\ 
    &= \int \int \dx \dy \frac{ \partial
    \Psi^{*} }{\partial q_{r}} \frac{ \partial \Psi }{ \partial q_{r} }
  \end{split}
  \intertext{and}
  \frac{ \partial \Psi{*} }{ \partial q_{r} } \frac{ \partial \Psi }{\partial
  q_{r} } &= \left( \frac{ \partial \left| \Psi \right| }{ \partial q_{r} }
  \right)^{2} + \left| \Psi \right|^{2} \left( \frac{ \partial S }{ \partial
  q_{r} }\right)^{2}.
 \end{align}
Since \(\left( \partial \left| \Psi \right| / \partial q_{r} \right)^{2} \geq
0\), we then have
  \begin{align}
    m^{2} \lang \left| \dot{q}_{r} \right|^{2} \rang_{eq} &\leq \lang
    \hat{\pi_{r}^{2}} \rang
    \intertext{and so}
    \lang \left| \delta q_{r} \left( t_{f} \right) \right| \rang_{eq} &\leq
    \frac{ 1 }{ m } \int_{t_i}^{t_f} \dt \sqrt{ \lang \hat{\pi_{r}^{2}} \rang}.
  \end{align}
We also have
  \begin{equation}
    \lang \hat{\pi_{r}^{2}} \rang < 2mW_{r}
  \end{equation}
where \(W_{r}\) denotes the \(x\)- or \(y\)- part of the mean Hamiltonian, with
\(W_{r} \propto M^{2}\). Hence,
  \begin{equation}
    \lang \left|\delta q_{r} \left( t_{f} \right) \right| \rang_{eq} <
    \int_{t_i}^{t_f} \dt \frac{1}{m} \sqrt{2mW_{r}}.
  \end{equation}
Since \(W_{r}\) is time-independent, and setting \(t_i = 0\) and \(t_f = t\)
we have
  \begin{equation}
    \lang \left| \delta q_{r} \left( t \right) \right| \rang_{eq} < t \sqrt{
    \frac{ 2 W_{r} }{ m }}.
  \end{equation}
Setting the right-hand side to be of order $L$, and noting that $\bar{E} \approx W_r$, then indeed yields the relaxation time in Eqn.~\ref{eqn:explanation}, whose inverse scaling with $M$ is in agreement with the numerical results presented above.

As mentioned, this derivation is based on that in Ref.~\onlinecite{valentini08}
but differs in some respects. The purpose of the analysis in
Ref.~\onlinecite{valentini08} was to derive a condition for the suppression of
relaxation in expanding space (here we are only concerned with static space) and
the condition for relaxation was that the mean displacement \(\delta q_{r}\) --
for field degrees of freedom in Fourier space -- should be comparable to (or
greater than) the quantum spread \(\Delta q_{r}\).  In the analysis above, the
only degree of freedom considered is the spatial displacement of a particle in
the potential well, the constraints of which slightly change the condition for
relaxation. Regardless of the spread in the wave function the particle cannot
move beyond the confines of the well, so the condition used for relaxation is
that the mean displacement of a particle is comparable to (or greater than) the
size of the potential well.

  \subsection{Relaxation time as a function of coarse-graining length}
  \label{sec:tau_vs_cg}

\begin{figure*}[p]
  \begin{center}
    \includegraphics[width=0.4\textwidth]
    {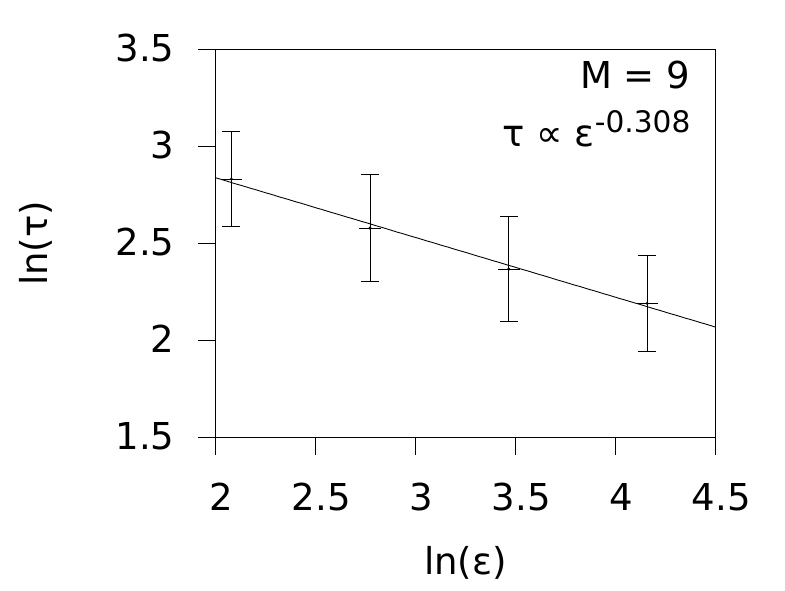}\hspace{0.1\textwidth}
    \includegraphics[width=0.4\textwidth]
    {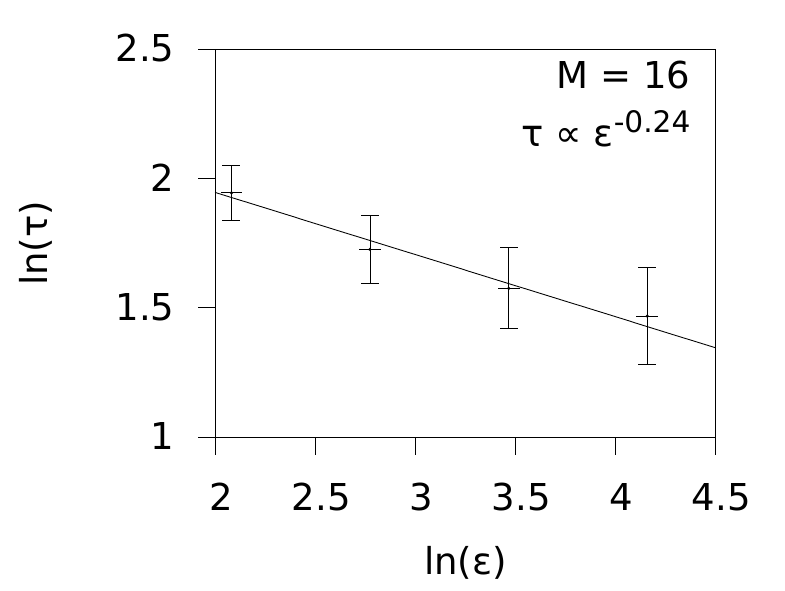}\\
    \includegraphics[width=0.4\textwidth]
    {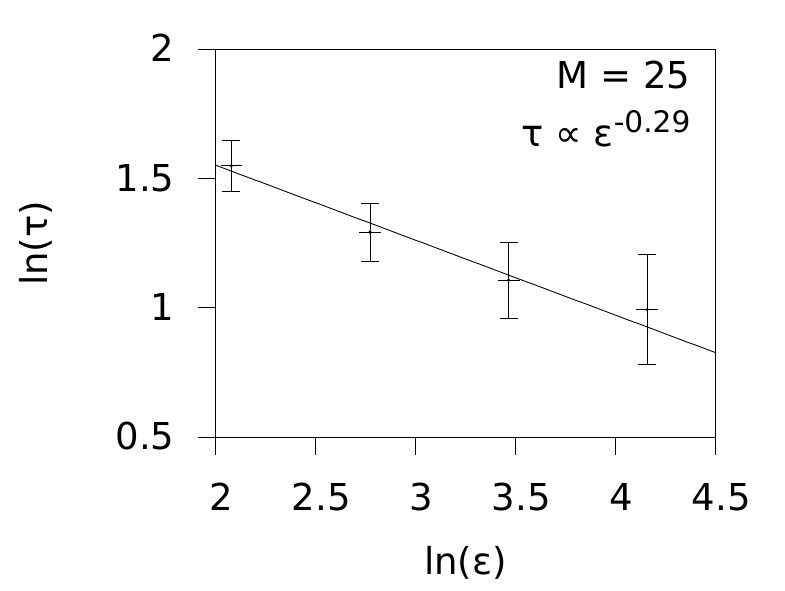}\hspace{0.1\textwidth}
    \includegraphics[width=0.4\textwidth]
    {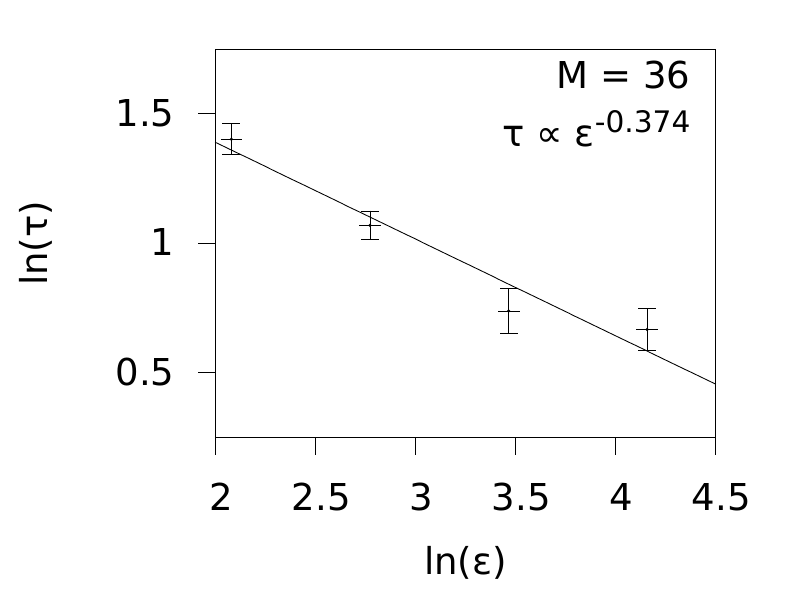}\\
    \includegraphics[width=0.4\textwidth]
    {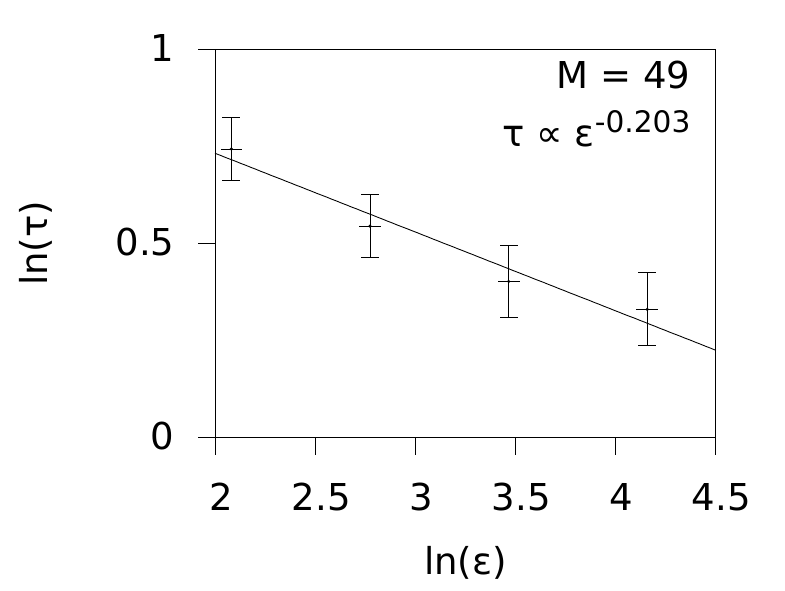}\hspace{0.1\textwidth}
    \includegraphics[width=0.4\textwidth]
    {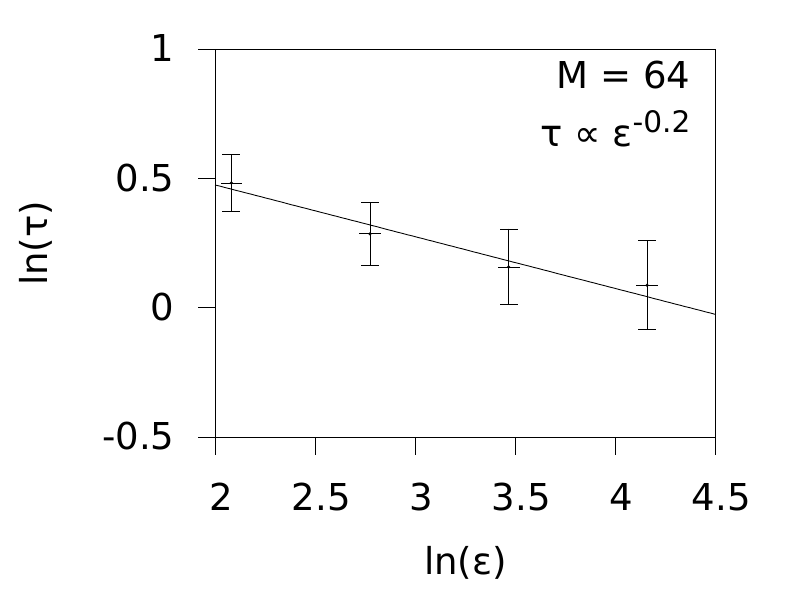}\\
    \caption{Graphs showing \(\ln \tau\) against \(\ln \varepsilon\). As in
    Figs.~\ref{fig:tau_vs_n_raw} and \ref{fig:tau_vs_n}, a logarithmic axis
    has been used, so as to identify a power law relationship. Also, as in
    Fig.~\ref{fig:tau_vs_n}, the datapoints with \(M = 4\) are excluded. The
    errors are estimated in the same way as those in Fig.~\ref{fig:tau_vs_n}.
    It is clear that the power law does not conform to
    Eqn.~\ref{eqn:prediction}, \(\tau \propto \varepsilon^{-1}\), nor
    does there seem to be a consistent power law. The
    straight line fit is not as convincing in this case as in the \(\tau\) vs
    \(M\) graphs, except for the case of \(M = 9\).}
    \label{fig:tau_vs_cg}
    \rule{0cm}{2cm}\\
  \end{center}
\end{figure*}

  \begin{SCtable*}[2.5][t]
    \begin{tabular}{|r|c|}
      \hline
      \hspace{.5cm}\(\phantom{\varepsilon}M\)\hspace{.5cm} & \(q\)\\
      \hline
      9 & \(-0.31 \pm 0.02\)\\
      \hline
      16 & \(-0.24 \pm 0.03\)\\
      \hline
      25 & \(-0.29 \pm 0.03\)\\
      \hline
      36 & \(-0.37 \pm 0.06\)\\
      \hline
      49 & \(-0.20 \pm 0.03\)\\
      \hline
      64 & \(-0.20 \pm 0.03\)\\
      \hline
    \end{tabular}
    \caption{Summary of results for a relationship between \(\tau\) and
    \(\varepsilon\) for various numbers of modes, shown graphically in
    Fig.~\ref{fig:tau_vs_cg}. The errors are estimated in the
    same way as the results in Table~\ref{tab:tau_vs_n}. These values are not
    compatible with Eqn.~\ref{eqn:prediction}, \(\tau
    \propto \varepsilon^{-1}\), nor do they appear to be compatible with a
    consistent power law.}
    \label{tab:tau_vs_cg}
  \end{SCtable*}

In Fig.~\ref{fig:tau_vs_cg}, the relaxation time $\tau$ is plotted against
coarse-graining length $\varepsilon$ for various $M$, again with logarithmic
axes and with the data for $M=4$ excluded for the same reason as before. The 
best-fit power law is shown on the graphs, and the results are summarized in
Table~\ref{tab:tau_vs_cg}. It is evident from these data that, although the
relaxation time evidently decreases with increasing \(\varepsilon\), the data do
not support Eqn.~\ref{eqn:prediction} nor are they
particularly suggestive of a constant power law.  A weak dependence of order
\(\tau \sim \varepsilon^{-^1/_4}\) is observed. Eqn.~\ref{eqn:prediction}
predicts a power law relation \(\tau \propto \varepsilon^{-1}\) while
Eqn.~\ref{eqn:explanation} predicts no dependence. 

As was the case with the $M$-dependence of $\tau$, the dependence on
\(\varepsilon\) is not in fact expected to take the form in
Eqn.~\ref{eqn:prediction}, for two reasons. First, the derivation of
Eqn.~\ref{eqn:prediction} is based on a definition of `relaxation time' that
applies only when $t \rightarrow 0$; it does not apply to the exponential tail
of $\cgh$, where most of the relaxation takes place. Second, the velocity field
can vary significantly over a coarse-graining cell, contrary to the assumption
made in the derivation of Eqn.~\ref{eqn:prediction}. Indeed the graphs in
Fig.~\ref{fig:tau_vs_cg} appear to show a systematic concave character rather
than a straight line,  apparently suggesting a small systematic deviation from a
power law model. The concave curvature is at least consistent with a power law
\(\tau \propto \varepsilon^{-1}\) in the limit \(\varepsilon \rightarrow 0\) (which must hold as we have noted), since
at some point to the left of the data in any of the figures the gradient ought
to approach (or pass through) \(-1\). A systematic study at smaller
coarse-graining lengths would however be rather difficult in terms of CPU time
because of the need for a significantly finer lattice and much greater overall
number of lattice points.

The lack of a dependence on \(\varepsilon\) in Eqn.~\ref{eqn:explanation} is not
surprising since the analysis in section~\ref{sec:explanation} and
\cite{valentini08} uses a different definition of the timescale. This definition
does not consider the \({\bar H}\)-function nor is there any necessity to
even mention coarse-graining. The weak dependence observed in the numerical
simulations should be interpreted as an effect outside the scope of this
prediction rather than one in conflict with it.

\section{Conclusions}
\label{sec:conclusions}

The numerical simulations performed in this work demonstrate clearly and
unequivocally the tendency for Born-rule distributions to arise spontaneously as
a consequence of ordinary pilot-wave dynamics, even for a system as simple as
the electron in a two-dimensional potential well. Contrary to popular belief
therefore, the Born rule does \emph{not} have to be introduced as a postulate of
non-relativistic quantum mechanics. What is the price paid for this? We must
suppose merely that particles have well-defined positions (and hence
trajectories) \emph{continuously} rather than only when a position measurement
is performed.

The main technical result of this work is the emergence of the  relationship
\(\tau \propto M^{-1}\) showing the dependence of the relaxation time on the
number of modes in a superposition. This result seems fairly robust under the
conditions studied here. In general terms, the faster relaxation times for
larger $M$ are due to the greater number of free-moving nodes in the pilot wave
which act as a source of vorticity and increase the chaotic nature of
trajectories. Our numerical result for the scaling conflicts with the
previous theoretical prediction, Eqn.~\ref{eqn:prediction}, but agrees with an alternative theoretical analysis presented here in
section~\ref{sec:explanation}. As we have discussed, the assumptions made in deriving Eqn.~\ref{eqn:prediction} were probably too restrictive for it to be of practical use. In particular, the defined timescale is relevant only close to $t = 0$, and does not apply to the exponential tail of the $\cgh$ function, where most of the relaxation takes place.

Our simulations reveal no well-defined scaling for the relaxation time as a
function of coarse-graining length \(\tau \left( \varepsilon \right)\), other
than a possible weak-dependence of the order of \(\tau \sim
\varepsilon^{-^{1}/_{4}}\). This also differs from Eqn.~\ref{eqn:prediction}
and  this is probably simply because the  coarse-graining cells are too large
for the derivation of Eqn.~\ref{eqn:prediction} to be valid. It is possible that
by decreasing $\varepsilon$ a behaviour conforming better to
Eqn.~\ref{eqn:prediction}( \(\tau \propto \varepsilon^{-1}\)) could be observed,
and it would be interesting to see at what length scale this begins to emerge.

Physically speaking our results suggest very short relaxation times with a range
of values observed for \(\tau\) between about 1 and 1000. Using natural units
\(c = \hbar =1\) and an electron with mass \(m = m_{e} = 1\) this corresponds to
relaxation times of the order of \(10^{-21}\)\textendash \(10^{-18}\)s. This is
consistent with our current understanding of quantum mechanics and modern
experimental investigations in which  no deviation from quantum equilibrium is
observed. If the initial state of the universe corresponded to a non-equilibrium
state, one might then assume that almost all deviations from the Born rule will
have been quickly washed out. However, it may be (see
Refs.~\onlinecite{valentini08, valentini10} and works cited therein) that relic
non-equilibrium from the early universe could be observed today, either directly
or by its imprint on the cosmic microwave background. In the future we
intend to modify the LOUIS program to deal with more realistic wave functions,
multi-particle systems, and expanding spaces with the intention of improving
some of the predictions outlined in Refs.~\onlinecite{valentini08, valentini10}.
More precise predictions for the effect on the CMB of non-equilibrium in the
early universe could lead to experimental tests of the de Broglie-Bohm
formulation of quantum mechanics.

How might these results generalize to more complex systems? In the original
version of the subquantum \emph{H}-theorem published by one of us (AV) in
1991~\cite{valentini91a, *valentini91b}, relaxation was considered for a
theoretical ensemble of complex \emph{N}-body systems. Once equilibrium is
reached for such an ensemble, it can be (and was) shown that extracting a single
particle from each system resulted in single-particle sub-ensembles that obey
the Born rule. The original view expressed in Ref.~\onlinecite{valentini91a,
*valentini91b} was that, realistically, relaxation would take place efficiently
only for many-body systems, and that the Born rule for single particles would be
derived by considering how these are extracted from more complex systems. But in
fact, in practice, there is an efficient relaxation even in simple
two-dimensional one-electron systems, and there appears to be no need to appeal
to a complex \emph{N}-body `parent system'.

Note that in a strict account of our world, say in the early universe, it could
well be that all degrees of freedom are entangled, so that there is no
actual ensemble of independent subsystems with the same wave function. However,
one can still talk about a theoretical ensemble of universes, each with the same
universal wave function, and consider the evolution of its distribution. (One
could also consider a mixed ensemble of universes, and apply our discussion to
each pure sub-ensemble.)

It is sometimes suggested that it is problematic to consider probabilities for
the `whole universe'. And yet, cosmologists are currently testing primordial
probabilities experimentally by measuring temperature anisotropies in the cosmic
microwave background. By making statistical assumptions about a theoretical
`ensemble of universes', cosmologists are able to test probabilities in the
early universe, such as those predicted by quantum field theory for vacuum
fluctuations during inflation. (For a detailed discussion of this in a de
Broglie-Bohm context, see Ref.~\cite{valentini10}.) One can question what the
ensemble of universes refers to. Is it a subjective probability distribution?
Or, is the universe we see in fact a member of a huge and perhaps infinite
ensemble, as is the case in theories of eternal inflation? Those are interesting
questions, but only tangentially related to the ongoing experimental tests. It is also important to bear in mind that there is much that is not known
about cosmology, so the treatment should be kept independent of cosmological
details as far as possible. What we do know is that all the particles we see
today are, or were, in complex superpositions (whether entangled with other
particles or not), and it appears clear from the simulations that such
superposition yields rapid relaxation - if it is rapid in two dimensions, one
would expect it to be even more rapid for 3$N$ dimensions.

\bibliographystyle{apsrev4-1.bst}
\bibliography{bibliography/pilot_waves}

\begin{thebibliography}{39}%
\makeatletter
\providecommand \@ifxundefined [1]{%
 \@ifx{#1\undefined}
}%
\providecommand \@ifnum [1]{%
 \ifnum #1\expandafter \@firstoftwo
 \else \expandafter \@secondoftwo
 \fi
}%
\providecommand \@ifx [1]{%
 \ifx #1\expandafter \@firstoftwo
 \else \expandafter \@secondoftwo
 \fi
}%
\providecommand \natexlab [1]{#1}%
\providecommand \enquote  [1]{``#1''}%
\providecommand \bibnamefont  [1]{#1}%
\providecommand \bibfnamefont [1]{#1}%
\providecommand \citenamefont [1]{#1}%
\providecommand \href@noop [0]{\@secondoftwo}%
\providecommand \href [0]{\begingroup \@sanitize@url \@href}%
\providecommand \@href[1]{\@@startlink{#1}\@@href}%
\providecommand \@@href[1]{\endgroup#1\@@endlink}%
\providecommand \@sanitize@url [0]{\catcode `\\12\catcode `\$12\catcode
  `\&12\catcode `\#12\catcode `\^12\catcode `\_12\catcode `\%12\relax}%
\providecommand \@@startlink[1]{}%
\providecommand \@@endlink[0]{}%
\providecommand \url  [0]{\begingroup\@sanitize@url \@url }%
\providecommand \@url [1]{\endgroup\@href {#1}{\urlprefix }}%
\providecommand \urlprefix  [0]{URL }%
\providecommand \Eprint [0]{\href }%
\@ifxundefined \urlstyle {%
  \providecommand \doi  [0]{\begingroup \@sanitize@url \@doi}%
  \providecommand \@doi [1]{\endgroup \@@startlink {\doibase
  #1}doi:\discretionary {}{}{}#1\@@endlink }%
}{%
  \providecommand \doi  [0]{doi:\discretionary{}{}{}\begingroup
  \urlstyle{rm}\Url }%
}%
\providecommand \doibase [0]{http://dx.doi.org/}%
\providecommand \Doi [0]{\begingroup \@sanitize@url \@Doi }%
\providecommand \@Doi  [1]{\endgroup\@@startlink{\doibase#1}\@@Doi}%
\providecommand \@@Doi [1]{#1\@@endlink}%
\providecommand \selectlanguage [0]{\@gobble}%
\providecommand \bibinfo  [0]{\@secondoftwo}%
\providecommand \bibfield  [0]{\@secondoftwo}%
\providecommand \translation [1]{[#1]}%
\providecommand \BibitemOpen [0]{}%
\providecommand \bibitemStop [0]{}%
\providecommand \bibitemNoStop [0]{.\EOS\space}%
\providecommand \EOS [0]{\spacefactor3000\relax}%
\providecommand \BibitemShut  [1]{\csname bibitem#1\endcsname}%
\bibitem [{\citenamefont {Saunders}\ \emph {et~al.}(2010)\citenamefont
  {Saunders}, \citenamefont {Barrett}, \citenamefont {Kent},\ and\
  \citenamefont {Wallace}}]{manyworldsbook}%
  \BibitemOpen
  \bibinfo {editor} {\bibfnamefont {S.}~\bibnamefont {Saunders}}, \bibinfo
  {editor} {\bibfnamefont {J.}~\bibnamefont {Barrett}}, \bibinfo {editor}
  {\bibfnamefont {A.}~\bibnamefont {Kent}}, \ and\ \bibinfo {editor}
  {\bibfnamefont {D.}~\bibnamefont {Wallace}},\ eds.,\ \href@noop {} {\emph
  {\bibinfo {title} {Many worlds? Everett, quantum theory, and reality}}}\
  (\bibinfo  {publisher} {Oxford University Press},\ \bibinfo {year}
  {2010})\BibitemShut {NoStop}%
\bibitem [{\citenamefont {Landsman}(2008)}]{compendium08}%
  \BibitemOpen
  \bibfield  {author} {\bibinfo {author} {\bibfnamefont {N.~P.}\ \bibnamefont
  {Landsman}},\ }in\ \href@noop {} {\emph {\bibinfo {booktitle} {Compendium of
  quantum physics}}},\ \bibinfo {editor} {edited by\ \bibinfo {editor}
  {\bibfnamefont {D.}~\bibnamefont {Greenberger}}, \bibinfo {editor}
  {\bibfnamefont {K.}~\bibnamefont {Hentschel}}, \ and\ \bibinfo {editor}
  {\bibfnamefont {F.}~\bibnamefont {Weinert}}}\ (\bibinfo  {publisher}
  {Springer},\ \bibinfo {year} {2008})\ pp.\ \bibinfo {pages}
  {64--69}\BibitemShut {NoStop}%
\bibitem [{\citenamefont {Born}(1926)}]{born26}%
  \BibitemOpen
  \bibfield  {author} {\bibinfo {author} {\bibfnamefont {M.}~\bibnamefont
  {Born}},\ }\href@noop {} {\bibfield  {journal} {\bibinfo  {journal} {Z.
  Phys.},\ }\textbf {\bibinfo {volume} {38}},\ \bibinfo {pages} {803} (\bibinfo
  {year} {1926})}\BibitemShut {NoStop}%
\bibitem [{\citenamefont {Heisenberg}(1927)}]{heisenberg27}%
  \BibitemOpen
  \bibfield  {author} {\bibinfo {author} {\bibfnamefont {W.}~\bibnamefont
  {Heisenberg}},\ }\href@noop {} {\bibfield  {journal} {\bibinfo  {journal} {Z.
  Phys.},\ }\textbf {\bibinfo {volume} {43}},\ \bibinfo {pages} {172} (\bibinfo
  {year} {1927})}\BibitemShut {NoStop}%
\bibitem [{\citenamefont {Bacciagaluppi}\ and\ \citenamefont
  {Valentini}(2009)}]{valentini_book}%
  \BibitemOpen
  \bibfield  {author} {\bibinfo {author} {\bibfnamefont {G.}~\bibnamefont
  {Bacciagaluppi}}\ and\ \bibinfo {author} {\bibfnamefont {A.}~\bibnamefont
  {Valentini}},\ }\href@noop {} {\emph {\bibinfo {title} {Quantum theory at the
  crossroads: reconsidering the 1927 {S}olvay conference}}}\ (\bibinfo
  {publisher} {Cambridge University Press},\ \bibinfo {year}
  {2009})\BibitemShut {NoStop}%
\bibitem [{\citenamefont {de~Broglie}(1928)}]{debroglie28}%
  \BibitemOpen
  \bibfield  {author} {\bibinfo {author} {\bibfnamefont {L.}~\bibnamefont
  {de~Broglie}},\ }in\ \href@noop {} {\emph {\bibinfo {booktitle} {Electrons et
  photons: rapports et discussions du cinquieme conseil de physique}}}\
  (\bibinfo  {publisher} {Gauthier-Villars, Paris},\ \bibinfo {year} {1928})\
  p.\ \bibinfo {pages} {105},\ \bibinfo {note} {{E}nglish translation in
  Ref.~\onlinecite{valentini_book}}\BibitemShut {NoStop}%
\bibitem [{\citenamefont {Bohm}(1952){\natexlab{a}}}]{bohm52a}%
  \BibitemOpen
  \bibfield  {author} {\bibinfo {author} {\bibfnamefont {D.}~\bibnamefont
  {Bohm}},\ }\href@noop {} {\bibfield  {journal} {\bibinfo  {journal} {Phys.
  Rev.},\ }\textbf {\bibinfo {volume} {85}},\ \bibinfo {pages} {166} (\bibinfo
  {year} {1952}{\natexlab{a}})}\BibitemShut {NoStop}%
\bibitem [{\citenamefont {Bohm}(1952){\natexlab{b}}}]{bohm52b}%
  \BibitemOpen
  \bibfield  {author} {\bibinfo {author} {\bibfnamefont {D.}~\bibnamefont
  {Bohm}},\ }\href@noop {} {\bibfield  {journal} {\bibinfo  {journal} {Phys.
  Rev.},\ }\textbf {\bibinfo {volume} {85}},\ \bibinfo {pages} {180} (\bibinfo
  {year} {1952}{\natexlab{b}})}\BibitemShut {NoStop}%
\bibitem [{\citenamefont {Bohm}\ and\ \citenamefont {Hiley}(1993)}]{bohm_book}%
  \BibitemOpen
  \bibfield  {author} {\bibinfo {author} {\bibfnamefont {D.}~\bibnamefont
  {Bohm}}\ and\ \bibinfo {author} {\bibfnamefont {B.~J.}\ \bibnamefont
  {Hiley}},\ }\href@noop {} {\emph {\bibinfo {title} {The undivided universe:
  an ontological interpretation of quantum theory}}}\ (\bibinfo  {publisher}
  {Routledge},\ \bibinfo {year} {1993})\BibitemShut {NoStop}%
\bibitem [{\citenamefont {Holland}(1993)}]{holland_book}%
  \BibitemOpen
  \bibfield  {author} {\bibinfo {author} {\bibfnamefont {P.~R.}\ \bibnamefont
  {Holland}},\ }\href@noop {} {\emph {\bibinfo {title} {The quantum theory of
  motion. An account of the de {B}roglie-{B}ohm causal interpretation of
  quantum mechanics}}}\ (\bibinfo  {publisher} {Cambridge University Press},\
  \bibinfo {year} {1993})\BibitemShut {NoStop}%
\bibitem [{\citenamefont {D{\"u}rr}\ and\ \citenamefont
  {Teufel}(2009)}]{durr_book}%
  \BibitemOpen
  \bibfield  {author} {\bibinfo {author} {\bibfnamefont {D.}~\bibnamefont
  {D{\"u}rr}}\ and\ \bibinfo {author} {\bibfnamefont {S.}~\bibnamefont
  {Teufel}},\ }\href@noop {} {\emph {\bibinfo {title} {Bohmian mechanics: the
  physics and mathematics of quantum theory}}}\ (\bibinfo  {publisher}
  {Springer},\ \bibinfo {year} {2009})\BibitemShut {NoStop}%
\bibitem [{\citenamefont {Riggs}(2009)}]{riggs_book}%
  \BibitemOpen
  \bibfield  {author} {\bibinfo {author} {\bibfnamefont {P.}~\bibnamefont
  {Riggs}},\ }\href@noop {} {\emph {\bibinfo {title} {Quantum causality:
  conceptual issues in the causal theory of quantum mechanics}}}\ (\bibinfo
  {publisher} {Springer},\ \bibinfo {year} {2009})\BibitemShut {NoStop}%
\bibitem [{\citenamefont {Valentini}(2009)}]{valentini_physics_world}%
  \BibitemOpen
  \bibfield  {author} {\bibinfo {author} {\bibfnamefont {A.}~\bibnamefont
  {Valentini}},\ }\href@noop {} {\enquote {\bibinfo {title} {Beyond the
  quantum},}\ }\bibinfo {howpublished} {Physics World, pp. 32-37} (\bibinfo
  {year} {November 2009})\BibitemShut {NoStop}%
\bibitem [{\citenamefont {Towler}(2009)}]{gradcourse}%
  \BibitemOpen
  \bibfield  {author} {\bibinfo {author} {\bibfnamefont {M.~D.}\ \bibnamefont
  {Towler}},\ }\href@noop {} {\enquote {\bibinfo {title} {Pilot waves,
  {B}ohmian metaphysics and the foundations of quantum mechanics},}\ }\bibinfo
  {howpublished} {Cambridge University graduate lecture course -
  \url{http://www.tcm.phy.cam.ac.uk/~mdt26/pilot_waves.html}} (\bibinfo {year}
  {2009})\BibitemShut {NoStop}%
\bibitem [{\citenamefont {Cushing}(1994)}]{cushing_book}%
  \BibitemOpen
  \bibfield  {author} {\bibinfo {author} {\bibfnamefont {J.~T.}\ \bibnamefont
  {Cushing}},\ }\href@noop {} {\emph {\bibinfo {title} {Quantum mechanics:
  historical contingency and the {C}openhagen hegemony}}}\ (\bibinfo
  {publisher} {University of Chicago Press},\ \bibinfo {year}
  {1994})\BibitemShut {NoStop}%
\bibitem [{\citenamefont {Towler}\ and\ \citenamefont
  {Valentini}(2010)}]{tticonf}%
  \BibitemOpen
  \bibfield  {author} {\bibinfo {author} {\bibfnamefont {M.~D.}\ \bibnamefont
  {Towler}}\ and\ \bibinfo {author} {\bibfnamefont {A.}~\bibnamefont
  {Valentini}},\ }\href@noop {} {\enquote {\bibinfo {title} {21st-century
  directions in de {B}roglie-{B}ohm theory and beyond},}\ }\bibinfo
  {howpublished} {\url{http://www.vallico.net/tti/deBB_10/conference.html}}
  (\bibinfo {year} {2010})\BibitemShut {NoStop}%
\bibitem [{\citenamefont {Valentini}(1991){\natexlab{a}}}]{valentini91a}%
  \BibitemOpen
  \bibfield  {author} {\bibinfo {author} {\bibfnamefont {A.}~\bibnamefont
  {Valentini}},\ }\href@noop {} {\bibfield  {journal} {\bibinfo  {journal}
  {Phys. Lett. A},\ }\textbf {\bibinfo {volume} {156}},\ \bibinfo {pages} {5}
  (\bibinfo {year} {1991}{\natexlab{a}})}\BibitemShut {NoStop}%
\bibitem [{\citenamefont {Valentini}(1991){\natexlab{b}}}]{valentini91b}%
  \BibitemOpen
  \bibfield  {author} {\bibinfo {author} {\bibfnamefont {A.}~\bibnamefont
  {Valentini}},\ }\href@noop {} {\bibfield  {journal} {\bibinfo  {journal}
  {Phys. Lett. A},\ }\textbf {\bibinfo {volume} {158}},\ \bibinfo {pages} {1}
  (\bibinfo {year} {1991}{\natexlab{b}})}\BibitemShut {NoStop}%
\bibitem [{\citenamefont {Valentini}(2002){\natexlab{a}}}]{valentini02}%
  \BibitemOpen
  \bibfield  {author} {\bibinfo {author} {\bibfnamefont {A.}~\bibnamefont
  {Valentini}},\ }\href@noop {} {\bibfield  {journal} {\bibinfo  {journal}
  {Pramana - J. Phys.},\ }\textbf {\bibinfo {volume} {59}},\ \bibinfo {pages}
  {269} (\bibinfo {year} {2002}{\natexlab{a}})}\BibitemShut {NoStop}%
\bibitem [{\citenamefont {Valentini}(2007)}]{valentini07}%
  \BibitemOpen
  \bibfield  {author} {\bibinfo {author} {\bibfnamefont {A.}~\bibnamefont
  {Valentini}},\ }\href@noop {} {\bibfield  {journal} {\bibinfo  {journal} {J.
  Phys. A},\ }\textbf {\bibinfo {volume} {40}},\ \bibinfo {pages} {3285}
  (\bibinfo {year} {2007})}\BibitemShut {NoStop}%
\bibitem [{\citenamefont {Valentini}(2010)}]{valentini10}%
  \BibitemOpen
  \bibfield  {author} {\bibinfo {author} {\bibfnamefont {A.}~\bibnamefont
  {Valentini}},\ }\href@noop {} {\bibfield  {journal} {\bibinfo  {journal}
  {Phys. Rev. D},\ }\textbf {\bibinfo {volume} {82}},\ \bibinfo {pages}
  {063513} (\bibinfo {year} {2010})}\BibitemShut {NoStop}%
\bibitem [{\citenamefont {Valentini}(1992)}]{valentini_thesis}%
  \BibitemOpen
  \bibfield  {author} {\bibinfo {author} {\bibfnamefont {A.}~\bibnamefont
  {Valentini}},\ }\href@noop {} {\enquote {\bibinfo {title} {On the pilot-wave
  theory of classical, quantum, and subquantum physics},}\ }\bibinfo
  {howpublished} {Ph.D. thesis, ISAS, Trieste, Italy
  \url{http://www.sissa.it/ap/PhD/Theses/valentini.pdf}} (\bibinfo {year}
  {1992})\BibitemShut {NoStop}%
\bibitem [{\citenamefont {Valentini}(2001)}]{valentini01}%
  \BibitemOpen
  \bibfield  {author} {\bibinfo {author} {\bibfnamefont {A.}~\bibnamefont
  {Valentini}},\ }in\ \href@noop {} {\emph {\bibinfo {booktitle} {Chance in
  Physics: Foundations and Perspectives (Lecture Notes in Physics)}}},\ Vol.\
  \bibinfo {volume} {574)},\ \bibinfo {editor} {edited by\ \bibinfo {editor}
  {\bibfnamefont {J.}~\bibnamefont {Bricmont}} \emph {et~al.}}\ (\bibinfo
  {publisher} {Springer},\ \bibinfo {year} {2001})\ pp.\ \bibinfo {pages}
  {165--181}\BibitemShut {NoStop}%
\bibitem [{\citenamefont {Valentini}\ and\ \citenamefont
  {Westman}(2005)}]{valentini05}%
  \BibitemOpen
  \bibfield  {author} {\bibinfo {author} {\bibfnamefont {A.}~\bibnamefont
  {Valentini}}\ and\ \bibinfo {author} {\bibfnamefont {H.}~\bibnamefont
  {Westman}},\ }\href@noop {} {\bibfield  {journal} {\bibinfo  {journal} {Proc.
  R. Soc. A},\ }\textbf {\bibinfo {volume} {461}},\ \bibinfo {pages} {253}
  (\bibinfo {year} {2005})}\BibitemShut {NoStop}%
\bibitem [{\citenamefont {Colin}\ and\ \citenamefont
  {Struyve}(2010)}]{colin09}%
  \BibitemOpen
  \bibfield  {author} {\bibinfo {author} {\bibfnamefont {S.}~\bibnamefont
  {Colin}}\ and\ \bibinfo {author} {\bibfnamefont {W.}~\bibnamefont
  {Struyve}},\ }\href@noop {} {\bibfield  {journal} {\bibinfo  {journal} {New
  Journal of Physics},\ }\textbf {\bibinfo {volume} {12}},\ \bibinfo {pages}
  {043008} (\bibinfo {year} {2010})}\BibitemShut {NoStop}%
\bibitem [{\citenamefont {Towler}(2010)}]{louis}%
  \BibitemOpen
  \bibfield  {author} {\bibinfo {author} {\bibfnamefont {M.~D.}\ \bibnamefont
  {Towler}},\ }\href@noop {} {\emph {\bibinfo {title} {\emph{The}
  \textsc{LOUIS} \emph{computer code}}}}\ (\bibinfo  {publisher} {University of
  Cambridge},\ \bibinfo {year} {2010})\BibitemShut {NoStop}%
\bibitem [{Note1()}]{Note1}%
  \BibitemOpen
  \bibinfo {note} {In the high-energy domain, pilot-wave theory for bosons
  usually takes the form of a field theory, while for fermions the best model
  invokes the Dirac sea. These models have a fundamental preferred rest frame,
  with an effective Lorentz invariance emerging only in equilibrium. For recent
  progress see in particular Refs.~\protect \rev@citealp {colin03,
  colinstruyve07, struyve10}.}\BibitemShut {Stop}%
\bibitem [{\citenamefont {Colin}\ \emph {et~al.}(2011)\citenamefont {Colin},
  \citenamefont {Struyve},\ and\ \citenamefont {Valentini}}]{bohm_instability}%
  \BibitemOpen
  \bibfield  {author} {\bibinfo {author} {\bibfnamefont {S.}~\bibnamefont
  {Colin}}, \bibinfo {author} {\bibfnamefont {W.}~\bibnamefont {Struyve}}, \
  and\ \bibinfo {author} {\bibfnamefont {A.}~\bibnamefont {Valentini}},\
  }\href@noop {} {\enquote {\bibinfo {title} {Instability of {B}ohm's
  dynamics},}\ } (\bibinfo {year} {2011}),\ \bibinfo {note} {in
  preparation}\BibitemShut {NoStop}%
\bibitem [{\citenamefont {Timko}\ and\ \citenamefont {Vrscay}(2009)}]{timko09}%
  \BibitemOpen
  \bibfield  {author} {\bibinfo {author} {\bibfnamefont {J.~A.}\ \bibnamefont
  {Timko}}\ and\ \bibinfo {author} {\bibfnamefont {E.~R.}\ \bibnamefont
  {Vrscay}},\ }\href@noop {} {\bibfield  {journal} {\bibinfo  {journal} {Found.
  Phys.},\ }\textbf {\bibinfo {volume} {39}},\ \bibinfo {pages} {1055}
  (\bibinfo {year} {2009})}\BibitemShut {NoStop}%
\bibitem [{\citenamefont {Bennett}(2010)}]{bennett10}%
  \BibitemOpen
  \bibfield  {author} {\bibinfo {author} {\bibfnamefont {A.~F.}\ \bibnamefont
  {Bennett}},\ }\href@noop {} {\bibfield  {journal} {\bibinfo  {journal} {J.
  Phys. A: Math. Theor.},\ }\textbf {\bibinfo {volume} {43}},\ \bibinfo {pages}
  {195304} (\bibinfo {year} {2010})}\BibitemShut {NoStop}%
\bibitem [{Note2()}]{Note2}%
  \BibitemOpen
  \bibinfo {note} {It should be noted, however, that in the code used by
  Valentini and Westman, the signs of the initial phases in the wave function
  were chosen with the opposite convention to that given in their text (where
  the latter agrees with Eqn.~\ref {eqn:wavefunc} above). Furthermore, in their
  plots of density functions, the labels on the $x$- and $y$-axes were
  inadvertently exchanged.}\BibitemShut {Stop}%
\bibitem [{\citenamefont {Valentini}(2002){\natexlab{b}}}]{valentini02b}%
  \BibitemOpen
  \bibfield  {author} {\bibinfo {author} {\bibfnamefont {A.}~\bibnamefont
  {Valentini}},\ }\href@noop {} {\bibfield  {journal} {\bibinfo  {journal}
  {Phys. Lett. A},\ }\textbf {\bibinfo {volume} {297}},\ \bibinfo {pages} {273}
  (\bibinfo {year} {2002}{\natexlab{b}})}\BibitemShut {NoStop}%
\bibitem [{Note3()}]{Note3}%
  \BibitemOpen
  \bibinfo {note} {Assuming $\rho $ and $\left | \Psi \right |^{2}$ are
  normalized then $\DOTSI \intop \ilimits@ (\rho - \left | \Psi \right
  |^{2})\protect \tmspace +\thinmuskip {.1667em}\protect \mathrm {d}\protect
  \mathbf {x}= 0$. Since $\rho \protect \qopname \relax o{ln}\left (\protect
  \frac {\rho }{\left | \Psi \right |^{2}} \right ) \geq \rho - \left | \Psi
  \right |^{2}$ for any value of $\rho $ and $\left | \Psi \right |^{2}$, with
  equality only when $\rho = \left | \Psi \right |^{2}$, then it is clear that
  $H = \DOTSI \intop \ilimits@ \rho \protect \qopname \relax o{ln}\left (
  \protect \frac {\rho }{\left | \Psi \right |^{2}} \right ) \protect \tmspace
  +\thinmuskip {.1667em}\protect \mathrm {d}\protect \mathbf {x}\geq 0$. Thus
  $H$ is bounded from below by zero.}\BibitemShut {Stop}%
\bibitem [{\citenamefont {Press}\ \emph {et~al.}(1992)\citenamefont {Press},
  \citenamefont {Flannery}, \citenamefont {Teukolsky},\ and\ \citenamefont
  {Vetterling}}]{numerical_recipes}%
  \BibitemOpen
  \bibfield  {author} {\bibinfo {author} {\bibfnamefont {W.~H.}\ \bibnamefont
  {Press}}, \bibinfo {author} {\bibfnamefont {B.~P.}\ \bibnamefont {Flannery}},
  \bibinfo {author} {\bibfnamefont {S.~A.}\ \bibnamefont {Teukolsky}}, \ and\
  \bibinfo {author} {\bibfnamefont {W.~T.}\ \bibnamefont {Vetterling}},\
  }\href@noop {} {\emph {\bibinfo {title} {Numerical Recipes in Fortran}}}\
  (\bibinfo  {publisher} {Cambridge University Press},\ \bibinfo {year}
  {1992})\ \bibinfo {note} {sections 16.1 and 16.2}\BibitemShut {NoStop}%
\bibitem [{\citenamefont {Towler}\ \emph {et~al.}(2010)\citenamefont {Towler},
  \citenamefont {Russell},\ and\ \citenamefont {Valentini}}]{density_video}%
  \BibitemOpen
  \bibfield  {author} {\bibinfo {author} {\bibfnamefont {M.~D.}\ \bibnamefont
  {Towler}}, \bibinfo {author} {\bibfnamefont {N.~J.}\ \bibnamefont {Russell}},
  \ and\ \bibinfo {author} {\bibfnamefont {A.}~\bibnamefont {Valentini}},\
  }\href@noop {} {\enquote {\bibinfo {title} {Video: dynamical relaxation to
  quantum equilibrium of an electron in a 2d potential well},}\ }\bibinfo
  {howpublished} {\url{http://www.tcm.phy.cam.ac.uk/~mdt26/raw_movie.gif}}
  (\bibinfo {year} {2010})\BibitemShut {NoStop}%
\bibitem [{\citenamefont {Valentini}(2008)}]{valentini08}%
  \BibitemOpen
  \bibfield  {author} {\bibinfo {author} {\bibfnamefont {A.}~\bibnamefont
  {Valentini}},\ }\href@noop {} {\enquote {\bibinfo {title} {De
  {B}roglie-{B}ohm prediction of quantum violations for cosmological
  super-{H}ubble modes},}\ }\bibinfo {howpublished} {arXiv:0804.4656v1
  [hep-th]} (\bibinfo {year} {2008})\BibitemShut {NoStop}%
\bibitem [{\citenamefont {Colin}(2003)}]{colin03}%
  \BibitemOpen
  \bibfield  {author} {\bibinfo {author} {\bibfnamefont {S.}~\bibnamefont
  {Colin}},\ }\href@noop {} {\bibfield  {journal} {\bibinfo  {journal} {Phys.
  Lett. A},\ }\textbf {\bibinfo {volume} {317}},\ \bibinfo {pages} {349}
  (\bibinfo {year} {2003})}\BibitemShut {NoStop}%
\bibitem [{\citenamefont {Colin}\ and\ \citenamefont
  {Struyve}(2007)}]{colinstruyve07}%
  \BibitemOpen
  \bibfield  {author} {\bibinfo {author} {\bibfnamefont {S.}~\bibnamefont
  {Colin}}\ and\ \bibinfo {author} {\bibfnamefont {W.}~\bibnamefont
  {Struyve}},\ }\href@noop {} {\bibfield  {journal} {\bibinfo  {journal} {J.
  Phys. A: Math. Theor.},\ }\textbf {\bibinfo {volume} {40}},\ \bibinfo {pages}
  {7309} (\bibinfo {year} {2007})}\BibitemShut {NoStop}%
\bibitem [{\citenamefont {Struyve}(2010)}]{struyve10}%
  \BibitemOpen
  \bibfield  {author} {\bibinfo {author} {\bibfnamefont {W.}~\bibnamefont
  {Struyve}},\ }\href@noop {} {\bibfield  {journal} {\bibinfo  {journal} {Rep.
  Prog. Phys.},\ }\textbf {\bibinfo {volume} {73}},\ \bibinfo {pages} {106001}
  (\bibinfo {year} {2010})}\BibitemShut {NoStop}%
\end{thebibliography}%

\end{document}